\documentclass[aps,prd,twocolumn,groupedaddress]{revtex4}
\usepackage{graphicx}
\usepackage{amsmath,amssymb,amsfonts}
\usepackage[colorlinks,citecolor=blue,linkcolor=blue,urlcolor=blue]{hyperref}

\usepackage{physics} 
\usepackage[percent]{overpic} 

\usepackage[english]{babel}

\usepackage[autostyle, english = british]{csquotes}
\MakeOuterQuote{"}

\begin{document}

\title{Evaporating black-to-white hole}

\author{P. Martin-Dussaud}
\email{pmd@cpt.univ-mrs.fr}
\affiliation{Aix Marseille Univ, Universit\'e de Toulon, CNRS, CPT, Marseille, France}
\author{C. Rovelli}
\affiliation{Aix Marseille Univ, Universit\'e de Toulon, CNRS, CPT, Marseille, France}

\date{\small\today}

\begin{abstract}
\noindent
We construct and discuss the form of the (effective) spacetime geometry inside a black hole undergoing a quantum transition to a white hole, taking into account the back-reaction of the component of the Hawking radiation falling into the hole.  
\end{abstract}

\maketitle 

\section{Introduction}

What is the ultimate fate of a black hole?  Classical general relativity disregards all quantum effects and predicts that black holes live forever, only allowed to grow  and never to shrink.  Hawking's celebrated result  \cite{Hawking1974a} has shown that quantum field theory predicts that a black hole of mass $m$ can emit radiation (the so-called Hawking quanta), as if it was a black body of temperature 
\begin{equation}\label{Hawking temperature}
T = \frac{1}{8 \pi m},
\end{equation}
in Planck units $\hbar = G = c = k_B = 1$ (see for instance \cite{Wald1975} for a detailed derivation).  By energy conservation, we then deduce that the back-reaction of the quantum matter on the geometry must make the black hole slowly to shrink.  The end of this evaporation process is outside the domain of validity of quantum field theory on a given background, and is not yet clear.  

Hawking suggested that the evaporation can continue until the simple disappearance of the black hole.  If so, unitarity of quantum evolution can be violated \cite{Hawking1976}. In addition, if the number of possible internal quantum states of a black hole is finite and bounded by the exponential of the horizon area, as some believe, unitarity is already lost long before complete evaporation, because this number is insufficient to purify the Hawking radiation \cite{Page1993a}. This loss of unitarity goes under the name of the "information-loss paradox". Many ideas have been suggested to address it.  Among these:  information is simply lost in the process \cite{Unruh2017a}; no black hole ever forms (fuzzballs) \cite{Mathur2005}; long-lived remnants \cite{Aharonov1987a}; firewalls \cite{Almheiri2013}; nonviolent information transfer from black holes \cite{Giddings2013a}; information leaks to planckian degrees of freedom \cite{Perez2015}. 

A complete theory of quantum gravity should adjudicate the issue, give a definite prediction of the ultimate fate of black holes, and tell us how the "information-loss paradox" is actually solved in nature.  Loop Quantum Gravity is one of the current approaches to the quantisation of general relativity.  It suggests the following solution to the paradox: (i) the horizon's area bounds the number of states that are distinguishable from the exterior during a time scale of the order of the black hole lifetime, but not the number of internal quantum states of the black hole (distinguishable by local quantum field observables inside the hole) \cite{Rovelli2017a}, thus evading the firewall theorem, (ii) a black hole does not simply pop out of existence at the end of its evaporation. Rather, it tunnels into a long living white hole before full evaporation \cite{Bianchi2018}. 

This \textit{black-to-white hole transition} scenario, also called \textit{fireworks} \cite{Haggard2014b} or the \emph{Planck star} scenario \cite{Rovelli2014b,Rovelli2017} is made possible by the existence of a solution of the classical Einstein equations which is compatible with a black hole undergoing an instantaneous and \emph{local} quantum transition to a white hole \cite{Haggard2014b} and is supported by direct calculations based on loop quantum gravity describing both the sole transition at the singularity \cite{Ashtekar2018,Ashtekar2018a} and the transition including the horizon \cite{Christodoulou2016,Christodoulou2018}. The black-to-white hole transition solves the information-loss paradox, since it gives information the possibility to be stored inside the hole and released by the white hole.

Two variations of the scenario are discussed in the literature; they differ by the estimated value of the black hole lifetime $\tau$, which depends on the mass $m$ of the initial black hole, in Planck units:
\begin{enumerate}
\item $\tau \sim m^2$, the tunnelling takes place while the black hole is still macroscopic, and Hawking evaporation can be neglected \cite{Haggard2014b, Rovelli2018a};
\item $\tau \sim m^3$, the tunnelling takes place after Hawking evaporation has shrunk the black hole to a nearly Planckian mass \cite{Bianchi2018}.
\end{enumerate}
The second variation was introduced in \cite{Bianchi2018} where it was suggested that Planck-mass white holes, resulting from exploding Planck-mass black holes, may be nothing else but long-lived remnants. The stability of 
Planck-mass white holes is discussed in \cite{Rovelli2018c}.

In both case, the metric undergoes a quantum tunneling at the time of transition from black to white hole. Strictly speaking there is no classical metric always in place, like there isn't a physically defined trajectory for a particle tunnelling under a potential barrier.  In the case of the particle tunnelling under a potential barrier, it is nevertheless still possible to define an effective trajectory, by connecting the partial semiclassical trajectories of the particle before and after the tunnelling.  This effective trajectory of course violates the classical equations of motion during the tunnelling.  The tunnelling is therefore modelled by a simple violation of the equations of motion. In a similar fashion, the black to white transition can be modelled by a single classical geometry that violates the classical Einstein equations in compact spatial region during a short time. 

In this paper, we construct and discuss the form that this effective spacetime geometry can take. Steps in this direction were taken in \cite{Haggard2014b,DeLorenzo2016} and \cite{Rovelli2018a}, but a crucial element was not taken into account: the Hawking radiation and its back-reaction. Here we improve on the understanding of the physics of the black-to-white hole transition by discussing possible ways of modelling the Hawking radiation and its back-reaction. Note that investigations on the same questionsn, altough following a different path, have been pursued by James M. Bardeen in \cite{Bardeen2018}.

In section II, we recall the general strategy of the semi-classical Einstein equations, how it can be applied to black holes, and how it motivates the construction of Hiscock model for evaporating black holes. In section III, we propose a toy model for an evaporating black-to-white hole, which is then improved by a carefully study of the evolution of the ingoing Hawking quanta beyond the singularity. In section IV, we motivates another possible model describing the evolution of outgoing quanta, and compare it to the previous one. In the conclusion, we finally recall how this work is just a step towards a more complete LQG computation.

\section{Model of evaporating black hole}

{\bf The semi-classical Einstein equations}

The complete description of black hole evaporation should require full quantum gravity, but an approximation can be obtained by quantum field theory on curved spacetime, as was done in the original derivation by Hawking \cite{Hawking1974a}. The gravitational degrees of freedom are described classically, and quantum matter fields evolve over it. The limit of such an approach is the back-reaction: the Hawking quanta, created over a classical space-time, are expected to affect in return the metric of this space-time. The metric of space-time, initially given by a vacuum solution of the Einstein equations ($G_{\mu \nu}=0$) should be modified to take into account the matter content of Hawking quanta ($T_{\mu \nu} \neq 0$). If back-reaction is neglected, then the Einstein equations are violated.

A possible approach to the problem is to consider a classical gravitational field $g_{\mu \nu}$ coupled to quantized matter fields, via the semiclassical Einstein equations
\begin{equation}\label{semiclassical EE}
G_{\mu \nu}(g_{\mu \nu}) = \matrixel{\psi}{\hat T_{\mu \nu}(g_{\mu \nu}) }{\psi},
\end{equation}
where $G_{\mu \nu}$ is the usual Einstein tensor (function of the metric $g_{\mu \nu}$), $\ket{\psi}$ is a quantum state for the matter, and $\hat T_{\mu \nu}$ is the quantized energy-momentum tensor of the matter. Equation \eqref{semiclassical EE} was first introduced by M\o ller as a general tool for approaching quantum gravity \cite{Moller1962}. An idea to solve it would be to use an iterative self-consistent method: 
\begin{enumerate}
\setlength\itemsep{0 \baselineskip}
\item start from a classical background metric $g^0_{\mu \nu}$;
\item compute $\matrixel{\psi}{\hat T_{\mu \nu}(g^0_{\mu \nu}) }{\psi}$ using QFT in curved space-time;
\item find $g^1_{\mu \nu}$ such that $G_{\mu \nu}(g^1_{\mu \nu}) = \matrixel{\psi}{\hat T_{\mu \nu}(g^0_{\mu \nu}) }{\psi}$;
\item iterate the procedure to find $g^2_{\mu \nu}$;
\item go on until it converges to a self-consistent solution $g_{\mu \nu}^\infty$ satisfying equation \eqref{semiclassical EE}.
\end{enumerate}
A lot of work has been done to compute the expectation value $\langle \psi | \hat T_{\mu \nu}(g_{\mu \nu}) | \psi \rangle$, for various metrics $g_{\mu \nu}$ and states $\ket{\psi}$, but the pursuit of the iterative method happens to be very hard, even at the first round.

\vspace{1 \baselineskip}

\newpage

{\bf Application to black holes}

\begin{figure}[h]
	\includegraphics[width = .7 \columnwidth]{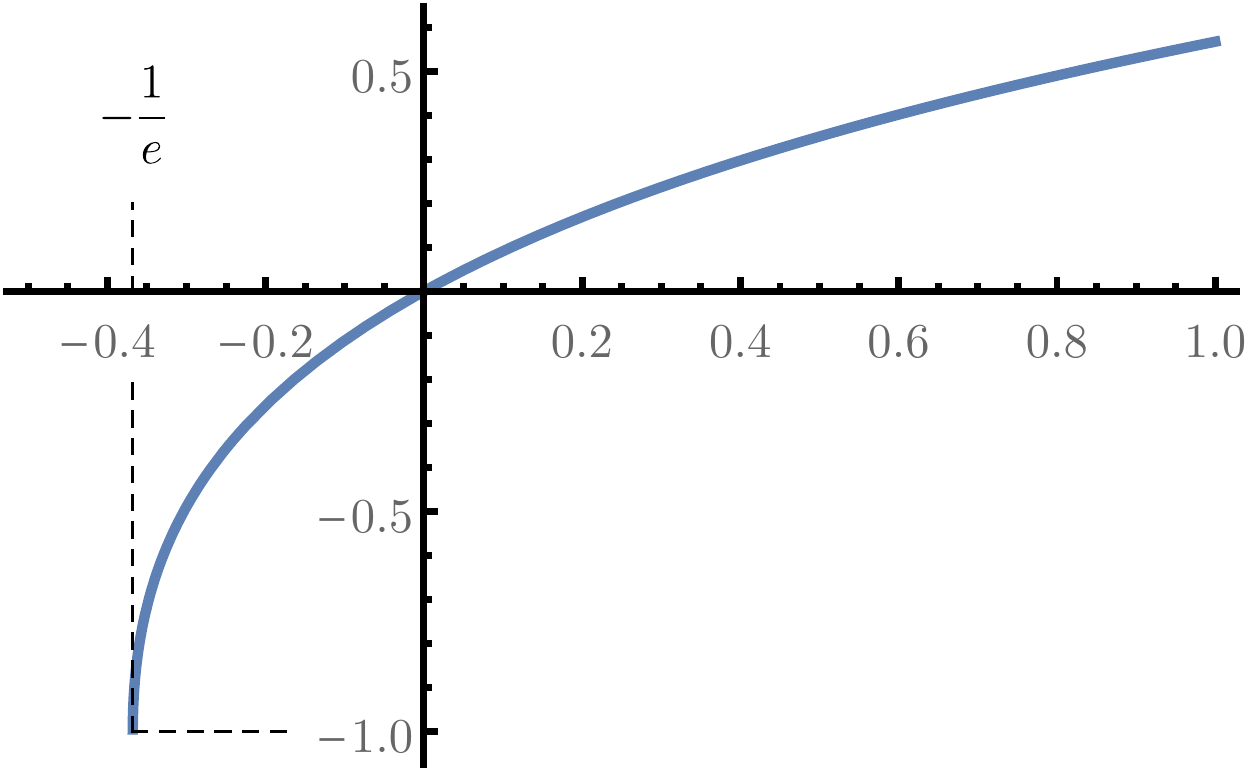}
	\caption{Graph of the upper branch of the Lambert $W$ function.}
	\label{Lambert}
\end{figure}

\begin{figure}[h]
\begin{overpic}[width =  0.9 \columnwidth]{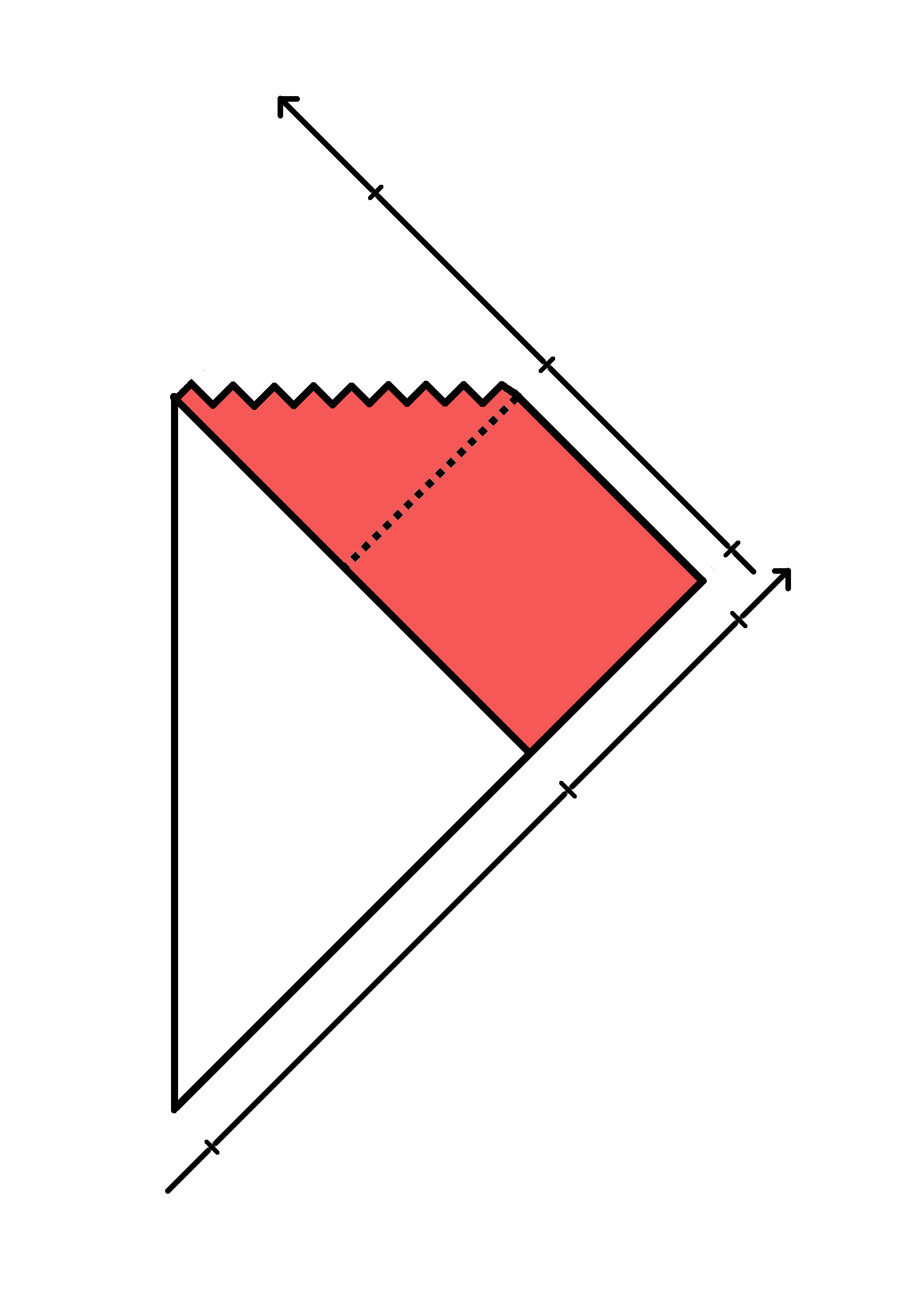}
\put (25,45) {$I$}
\put (35,55) {$IIa$}
\put (25,65) {$IIb$}
\put (15,92) {$U$}
\put (50,65) {$\mathcal{J}^+$}
\put (30,88) {$\frac{\pi}{2} - V_0$}
\put (44,74) {$0$}
\put (58,60) {$- \frac{\pi}{2}$}
\put (63,53) {$V$}
\put (58,50) {$\frac{\pi}{2}$}
\put (45,35) {$V_0 > 0$}
\put (18,8) {$2 V_0 - \pi $}
\end{overpic}
	\caption{Penrose diagram of a black hole formed by the collapse of a null shell. The event horizon is depicted with a dashed line. }
	\label{simple-black-hole}
\end{figure}

For a two-dimensional black hole formed by the collapse of a null shell, Hiscock was able to compute the expectation value $\langle in | \hat T_{\mu \nu} (g^0_{\mu \nu}) | in \rangle$ with a state $\ket{in}$ that matches the Minkowski vacuum inside the shell \cite{Hiscock1981}. The Penrose diagram of the model is shown on Figure \ref{simple-black-hole} with the metric $g^0_{\mu \nu}$ given by
\begin{align}
\label{simple-black-hole-metricI} (I)  &
\left[
\begin{array}{l}
\dd s^2 = - du dv + r^2 d\Omega^2 \\
r= \frac{1}{2} \left( v - u \right)
\end{array}
\right. \\
\label{simple-black-hole-metricIIa} (IIa) &
\left[
\begin{array}{l}
\dd s^2 = - \left(1-\frac{2m}{r}\right) du dv + r^2 d\Omega^2\\
r = 2m \left( 1+ W\left( e^{\frac{v-u}{4m} -1} \right) \right) 
\end{array}
\right. \\
 \label{simple-black-hole-metricIIb} (IIb)&
\left[
\begin{array}{l}
\dd s^2 = \left(1-\frac{2m}{r}\right) du dv + r^2 d\Omega^2\\
r = 2m \left( 1+ W\left( - e^{\frac{v+u}{4m} -1} \right) \right) 
\end{array}
\right. 
\end{align}
with $d\Omega^2 = d\theta^2 + \sin^2 \theta d\phi^2$ the usual metric of the unit sphere.
The map between the metric coordinates $(u,v)$ and the coordinates of the diagram $(U,V)$ is
\begin{align}
(I) &
\left[
\begin{array}{l}
u = v_0 - 4m \left( 1 + W ( - e^{-1} \tan V_0 \tan U ) \right) \\
v = v_0 - 4m \big( 1 + W ( - e^{-1} \tan V_0  \\
\hfill \quad \times \tan (V - 2 V_0 + \pi/2)  ) \big) \\
\hfill \text{with} \  v_0 \overset{\rm def}= 4m \log \tan V_0
\end{array}
\right. \\
\label{change coord simple IIa}
(IIa) &
\left[
\begin{array}{l}
u = - 4m \log (-\tan U) \\
v = 4 m \log \tan V 
\end{array}
\right. \\
\label{change coord simple IIb}
(IIb) &
\left[
\begin{array}{l}
u = 4m \log \tan U \\
v = 4 m \log \tan V
\end{array}
\right. 
\end{align}
The function $W$ is the upper branch of the Lambert $W$ function. It is an increasing function defined by the equation $x = W(x) e^{W(x)}$ and its graph is shown in Figure \ref{Lambert}.

In region $I$, $\langle in | \hat T_{\mu \nu} (g^0_{\mu \nu}) | in \rangle$, abbreviated $\expval{T_{\mu \nu}}$, vanishes everywhere, while in region $II$, the various components are given by
\begin{align}
&\expval{T_{uu}} = \frac{\hbar}{24 \pi} \left[ - \frac{m}{r^3} + \frac{3m^2}{2r^4} + \frac{m}{ r(u,v_0)^3} - \frac{3m^2}{2  r(u,v_0)^4} \right] \label{T_uu} \\
&\expval{T_{vv}} = \frac{\hbar}{24 \pi} \left[ - \frac{m}{r^3} + \frac{3m^2}{2r^4} \right] \label{T_vv}\\
&\expval{T_{uv}} = -  \frac{\hbar}{24 \pi} \left(1- \frac{2m}{r} \right) \frac{m}{r^3}.\label{T_uv}
\end{align}
Notice first that these formulae are valid both outside and inside the hole, although the coordinates $(u,v)$ map the two patches $IIa$ and $IIb$ in a different way (see equations \eqref{change coord simple IIa} and \eqref{change coord simple IIb}). Notice then that $\expval{T_{\mu \nu}}$ include both contributions from vacuum polarisation (the so-called Boulware state $\ket{B}$) and from Hawking quanta:
\begin{equation}
\expval{T_{\mu \nu}} = \matrixel{B}{T_{\mu \nu}}{B} + \matrixel{in}{:T_{\mu \nu}:}{in}.
\end{equation}
The Hawking flux contribution comes only from the normal ordered stress tensor, whose non-vanishing components are in the outgoing null direction \cite{Fabbri2005}:
\begin{equation}\label{Hawking flux}
\begin{split}
&\matrixel{in}{:T_{uu}:}{in} = \frac{\hbar}{24 \pi} \left[\frac{m}{ r(u,v_0)^3} - \frac{3m^2}{2  r(u,v_0)^4} \right] \\
&\matrixel{in}{:T_{vv}:}{in} = 0 \\
&\matrixel{in}{:T_{uv}:}{in} = 0. 
\end{split}
\end{equation}

Clearly $g^0_{\mu \nu}$ does not solve the semi-classical Einstein equations, since $G_{\mu \nu}(g^0_{\mu \nu}) = 0$, while in region $II$, $\langle in | \hat T_{\mu \nu} (g^0_{\mu \nu}) | in \rangle \neq 0$. So the idea of the iterative approach was to propose a corrected metric $g^1_{\mu \nu}$, that would ideally solve
\begin{equation}
G_{\mu \nu}(g^1_{\mu \nu}) = \langle in | \hat T_{\mu \nu} (g^0_{\mu \nu}) | in \rangle.
\end{equation} 
Unfortunately, solving this equation seems to be already too hard. Then, Hiscock suggested to \textit{guess} a metric $g^1_{\mu \nu}$, that would violate the semi-classical Einstein equations \textit{less} than the original background $g^0_{\mu \nu}$. This lead him to devise a model for an evaporating space-time that we now recall \cite{Hiscock1981a}.

\vspace{\baselineskip}

{\bf Hiscock model}

How to guess a corrected metric? We can take inspiration from the value of $\expval{T_{\mu \nu}}$ in some regions. In our case two regions are noticeable. First, along future null infinity, $\mathcal{J}^+$, the only non-vanishing component is
\begin{equation}
\expval{T_{uu}} = \frac{\hbar}{24 \pi} \left[\frac{m}{ r(u,v_0)^3} - \frac{3m^2}{2  r(u,v_0)^4} \right] .
\end{equation}
To understand intuitively what it means, suppose, in $2$-dimensional Minkowski space ($ds^2 = - du dv$), that the same kind of stress-energy tensor is due to isolated particles, i.e. $T_{\mu \nu} = \rho \, u_\mu u_\nu$, with $u^\mu$ the four-momentum. Then, if $T_{uu}$ is the only non-vanishing component, it means that $u_{\mu} \propto (1,0)$, in the $(\partial_u, \partial_v)$ basis, and so $u^\mu \propto (0,1)$, which mean particles are going away along the $v$ direction. Since $\expval{T_{uu}} > 0$ on $\mathcal{J}^+$, we have the picture of particles of positive energy reaching $\mathcal{J}^+$ along null geodesics directed by $\partial_v$. Indeed, the black hole evaporates. 

Secondly, along the horizon, $r=2m$, the only non-vanishing component is
\begin{equation}
\expval{T_{vv}} = - \frac{\hbar}{768 \pi m^2}.
\end{equation}
This time we can have the picture that particles of negative energy are leaving the horizon along null geodesics directed by $\partial_u$.

\begin{figure}[h]
\begin{overpic}[width =  0.9 \columnwidth]{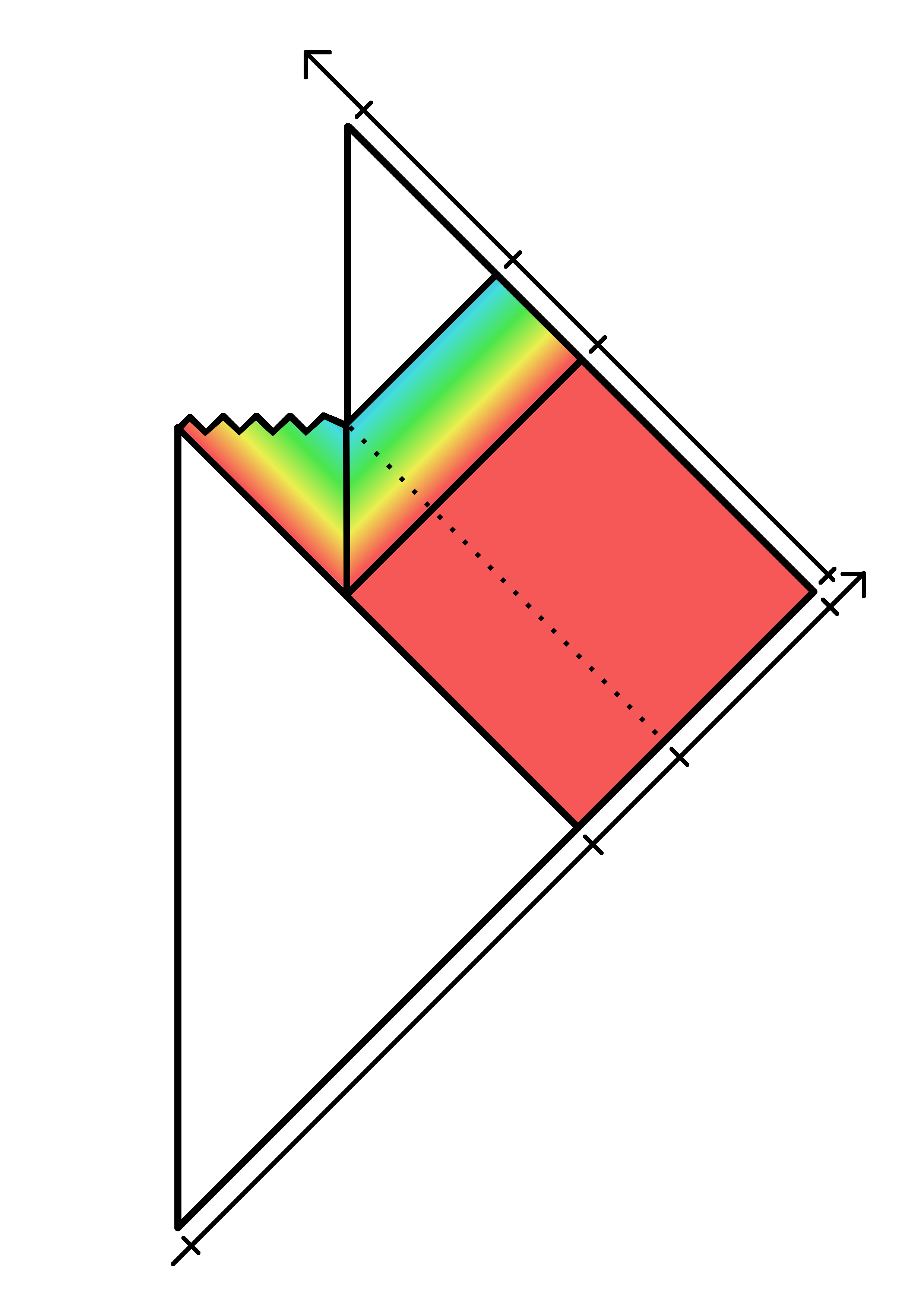}
\put (25,35) {$I$}
\put (45,55) {$II$}
\put (20,62) {$III$}
\put (33,67) {$IV$}
\put (30,78) {$V$}
\put (68,55) {$V$}
\put (65,50) {$\frac{\pi}{2}$}
\put (55,40) {$V_0 $}
\put (48,32) {$\frac{\pi}{4} $}
\put (18,3) {$\frac{\pi}{2} - 2 V_0 $}
\put (18,95) {$U$}
\put (30,93) {$0$}
\put (40,82) {$ V_0 - \frac{\pi}{2}$}
\put (48,75) {$- \frac{\pi}{4}$}
\put (61,60) {$- \frac{\pi}{2}$}
\end{overpic}
	\caption{Penrose diagram of Hiscock model. Everywhere the metric is locally that of Schwarzschild, characterised by a parameter of mass. Its value is represented by a color, from white (mass $0$, i.e. Minkoswki) to red (initial mass $m$), passing through a gradient ($M(u)$ or $N(v)$). The mass profile along $\mathcal{J}^+$ is shown on Figure \ref{mass-Hiscock}.}
	\label{HiscockII}
\end{figure}
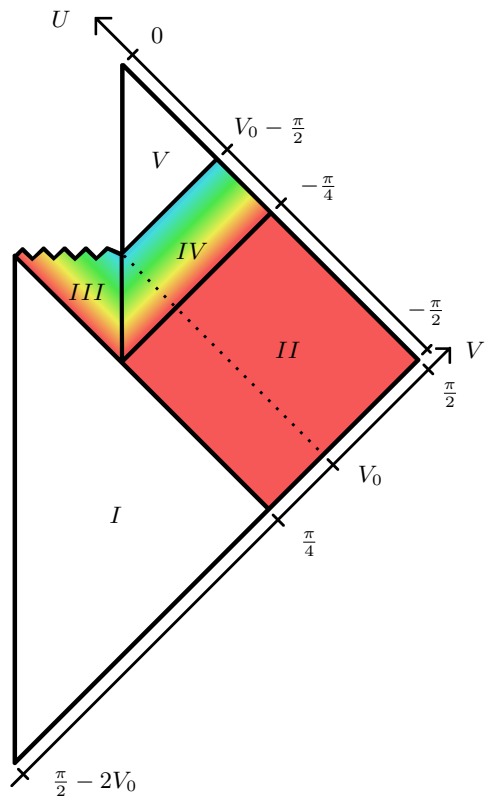

These two pictures motivate the model of Hiscock. It cleverly uses Vaidya-like metrics to represent the two fluxes of particles. It is made of five patches glued together as shown on Figure \ref{HiscockII}, and the metric (the "guessed" $g^1_{\mu \nu}$) is given by
\begin{align}
(I) \label{Hiscock metric I} &
\left[
\begin{array}{l}
\dd s^2 = - du dv + r^2 d\Omega^2 \\
r = \frac12 (v-u)
\end{array}
\right. \\
(II) &
\left[
\begin{array}{l}
\dd s^2 = - \left(1-\frac{2m}{r}\right) du dv + r^2 d\Omega^2 \\
r = 2m \left( 1+ W\left( e^{\frac{v-u}{4m} -1} \right) \right) 
\end{array}
\right. \\
(III) &
\left[
\begin{array}{l}
\dd s^2 = - \left(1-\frac{2N(v)}{r}\right) dv^2 + 2dv dr + r^2 d\Omega^2 
\end{array}
\right. \\
(IV) &
\left[
\begin{array}{l}
\dd s^2 = - \left(1-\frac{2M(u)}{r}\right) du^2 - 2du dr + r^2 d\Omega^2 
\end{array}
\right. \label{Hiscock metric IV} \\
(V) &
\left[
\begin{array}{l}
\dd s^2 = - du dv + r^2 d\Omega^2 \\
r = \frac12 (v-u)
\end{array}
\right. \label{Hiscock metric V}
\end{align}
The metric depends on the mass $m$ of the black hole at the beginning of the evaporation. It also makes use of two functions $M(u)$ and $N(v)$, which represent how the mass decreases with the evaporation. Their value matches along the boundary $III/IV$, which marks the \textit{apparent horizon}.

For completeness of the construction, we shall give the formulae that relates the coordinates $(U,V)$ of the Penrose diagram to the coordinates in which the metric of each patch is written. This is not fully done in the original paper of Hiscock \cite{Hiscock1981a}, but it is a necessary work to show that the Penrose diagram of Figure \ref{HiscockII} correctly represents a consistent space-time model. The scrupulous reader will find the equations in appendix \ref{details Hiscock}.

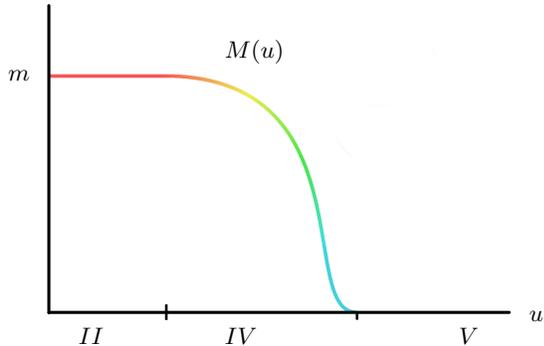
\begin{figure}[h]
\begin{overpic}[width =  0.9 \columnwidth]{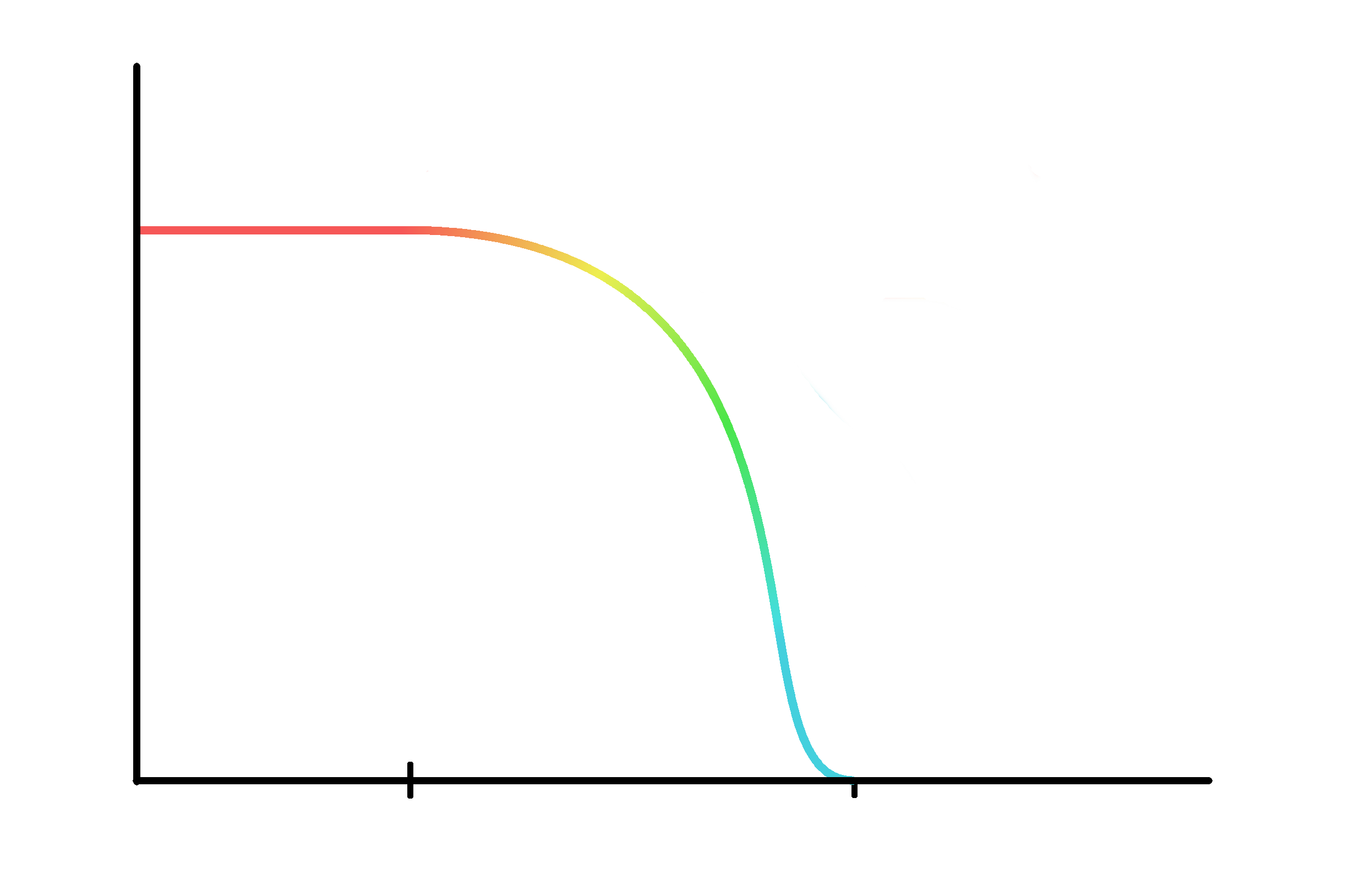}
\put (3,48) {$m$}
\put (92,7) {$u$}
\put (40,52) {$M(u)$}
\put (15,3) {$II$}
\put (40,3) {$IV$}
\put (80,3) {$V$}
\end{overpic}
	\caption{Bondi-Sachs mass function along $\mathcal{J}^+$ for Hiscock model.}
	\label{mass-Hiscock}
\end{figure}

How shall we choose the mass function $M(u)$ of the model? From Hawking's temperature formula \eqref{Hawking temperature}, the rate of mass loss was estimated by Page (see \cite{Page1976}) as 
\begin{equation}
\frac{\dd M}{\dd t} \propto - \frac{1}{M^2}.
\end{equation}
This suggests the behavior $M(u) \sim (u_0 - u)^{1/3}$, where $u_0$ is the retarded time at which the black hole faints, holding as long as the semi-classical approximation is valid. Nevertheless Hiscock shows that this behavior cannot hold until the end of the evaporation, and that a finite total amount of energy flux on $\mathcal{J}^+$ implies that
\begin{equation}
\lim_{M \to 0} \frac{\dd M}{\dd u} = 0.  
\end{equation} 
Therefore, Hiscock proposes a mass profile shown on Figure \ref{mass-Hiscock}.

From the perspective of an outside observer, Hiscock model seems to describe correctly the phenomenology expected at the first stages of the evaporation. The decreasing Bondi-Sachs mass $M(u)$ along $\mathcal{J}^+$ corresponds to an outgoing positive energy flux, due to Hawking radiation. According to this model, the black hole evaporates completely and space-time turns to Minkowski back again. From our perpspective this end scenario is more disputable for the persistence of the singularity than for a potential loss of unitarity. In the following section, we consider a possible white future to the singularity, which as a spin-off, gives a way to restore unitarity.

\section{Model of evaporating black-to-white hole (I)} \label{section model I}

In \cite{Haggard2014b}, a first space-time model was proposed to describe a quantum tunnelling from a black hole to a white hole, neglecting deliberately Hawking evaporation in the process. Then we proposed in \cite{Rovelli2018a} an alternative model where the geodesics can go accross the singularity continuously.  In \cite{Bianchi2018}, Bianchi et al. argued that a black-to-white hole transition would be much more probable at the end of Hawking evaporation, when typically the black hole has reached a  Planckian mass $m_1$. It is the goal of what follows to build an explicit model for an evaporating black-to-white hole.

\vspace{\baselineskip}
{\bf Toy model}

\begin{figure}[h]
\begin{overpic}[width =  0.9 \columnwidth]{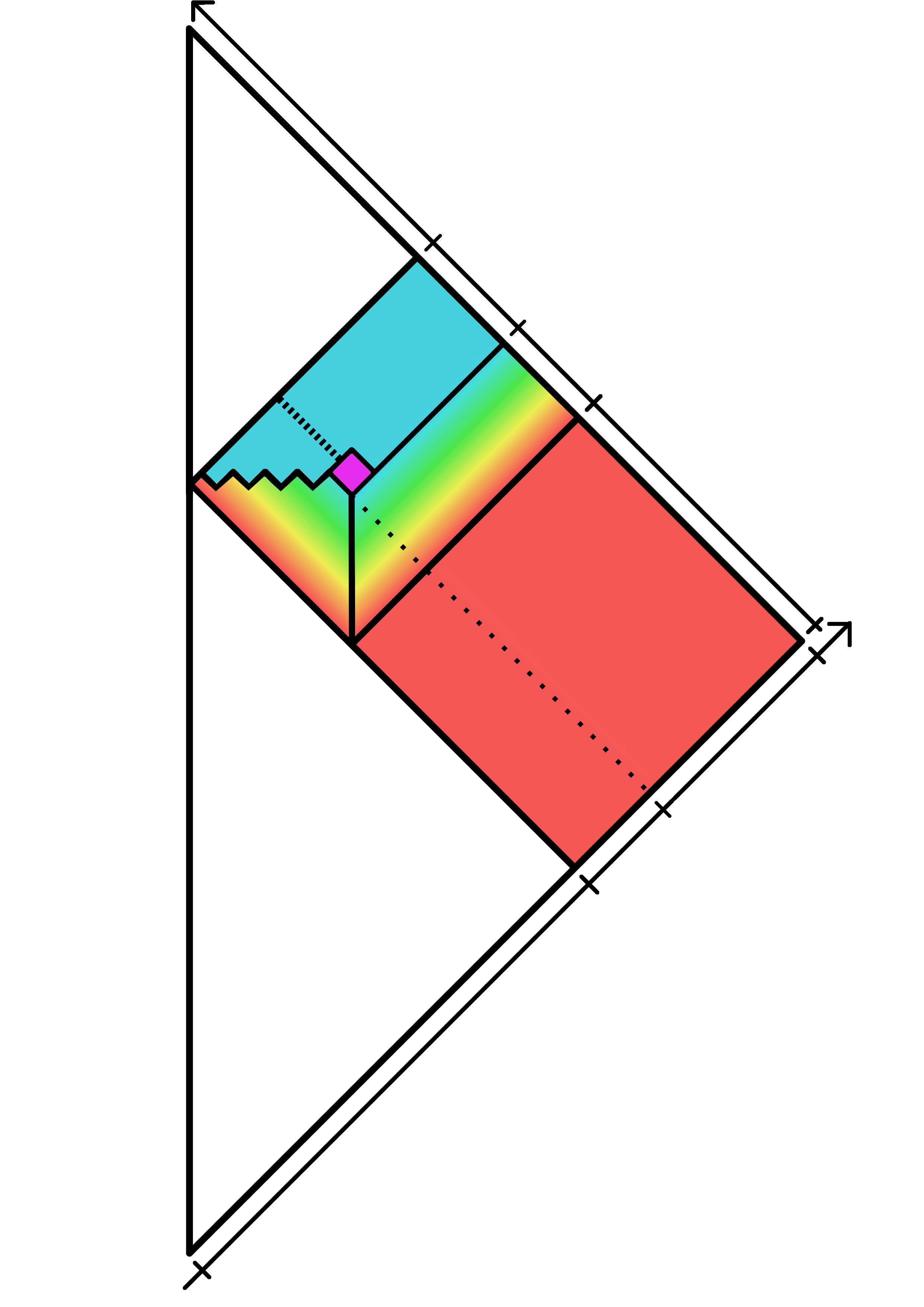}
\put (25,35) {$I$}
\put (45,50) {$II$}
\put (20,58) {$III$}
\put (33,65) {$IV$}
\put (30,70) {$V$}
\put (20,65) {$VI$}
\put (20,80) {$VII$}
\put (68,50) {$V$}
\put (63,45) {$\frac{\pi}{2}$}
\put (52,35) {$V_0-\varepsilon $}
\put (48,28) {$\frac{\pi}{4} $}
\put (18,0) {$\frac{\pi}{2} - 2 V_0 $}
\put (10,100) {$U$}
\put (34,84) {$ 2V_0 - \frac{3\pi}{4}$}
\put (40,76) {$ V_0 -\varepsilon- \frac{\pi}{2}$}
\put (48,70) {$- \frac{\pi}{4}$}
\put (61,55) {$- \frac{\pi}{2}$}
\end{overpic}
	\caption{Penrose diagram of a toy model of an evaporating black hole that turns into a white hole.}
	\label{model1}
\end{figure}
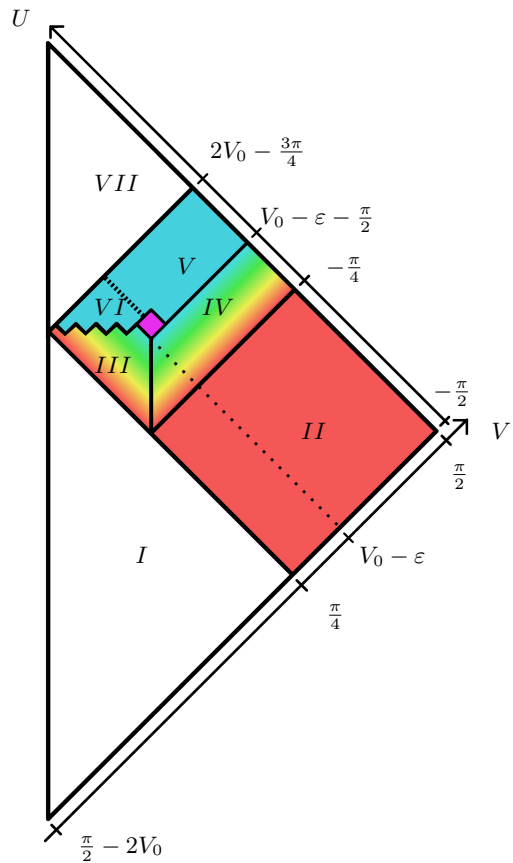

The initial idea is simple. Given the Hiscock model of an evaporating black hole, depicted on Figure \ref{HiscockII}, just glue a white hole above the singularity, with an outgoing bouncing null shell. This can be done easily provided that the Bondi-Sachs mass $M(u)$, observed on $\mathcal{J}^+$ in region $(IV)$, does not vanish completely but reaches a small positive value $m_1$. We obtain the Penrose diagram of Figure \ref{model1}.

In region $I, II, III, IV$ the metric is the same as the model of Hiscock (see equations (\ref{Hiscock metric I}-\ref{Hiscock metric IV}) and (\ref{HiscockII-Ia}-\ref{HiscockII-IV})). Elsewhere, the metric is given by
\begin{align}
(V) &
\left[
\begin{array}{l}
\dd s^2 = - \left(1-\frac{2m_1}{r}\right) du dv + r^2 d\Omega^2\\
r = 2m_1 \left( 1+ W\left( e^{\frac{v-u}{4m_1} -1} \right) \right) 
\end{array}
\right. \\
(VI) &
\left[
\begin{array}{l}
\dd s^2 = \left(1-\frac{2m_1}{r}\right) du dv + r^2 d\Omega^2\\
r = 2m_1 \left( 1+ W\left( - e^{- \frac{v+u}{4m_1} -1} \right) \right) 
\end{array}
\right. \\
(VII) &
\left[
\begin{array}{l}
\dd s^2 = - du dv + r^2 d\Omega^2 \\
r= \frac{1}{2} \left( v - u \right)
\end{array}
\right. 
\end{align}
$\varepsilon$ is defined so that the radius at the future endpoint of the apparent horizon is $2m_1$, i.e. $h(V_0-\varepsilon-\pi/2,V_0 - \varepsilon) = 2m_1$, where $h(U,V)$ is defined in appendix \ref{details Hiscock}. The details of the map between the coordinates $(u,v)$ of the metric and $(U,V)$ of the diagram are postponed to appendix \ref{detail toy model}.

The central purple diamond is a very small region of space-time. We have not given an explicit expression for the metric here but it would a priori be possible to construct one that matches the boundary conditions around. It is believed to be a region where quantum effects happen to enable the tunnelling to the white hole. Thus, it would be better described by a quantum geometry, instead of any effective classical metric. The Einstein equations are necessarily violated in this region since classical general relativity does not allow the black-to-white hole scenario. The novelty with respect to previous models like \cite{Rovelli2018a} is that the region is very small, typically Planckian.

\begin{figure}[h]
\begin{overpic}[width =  0.9 \columnwidth]{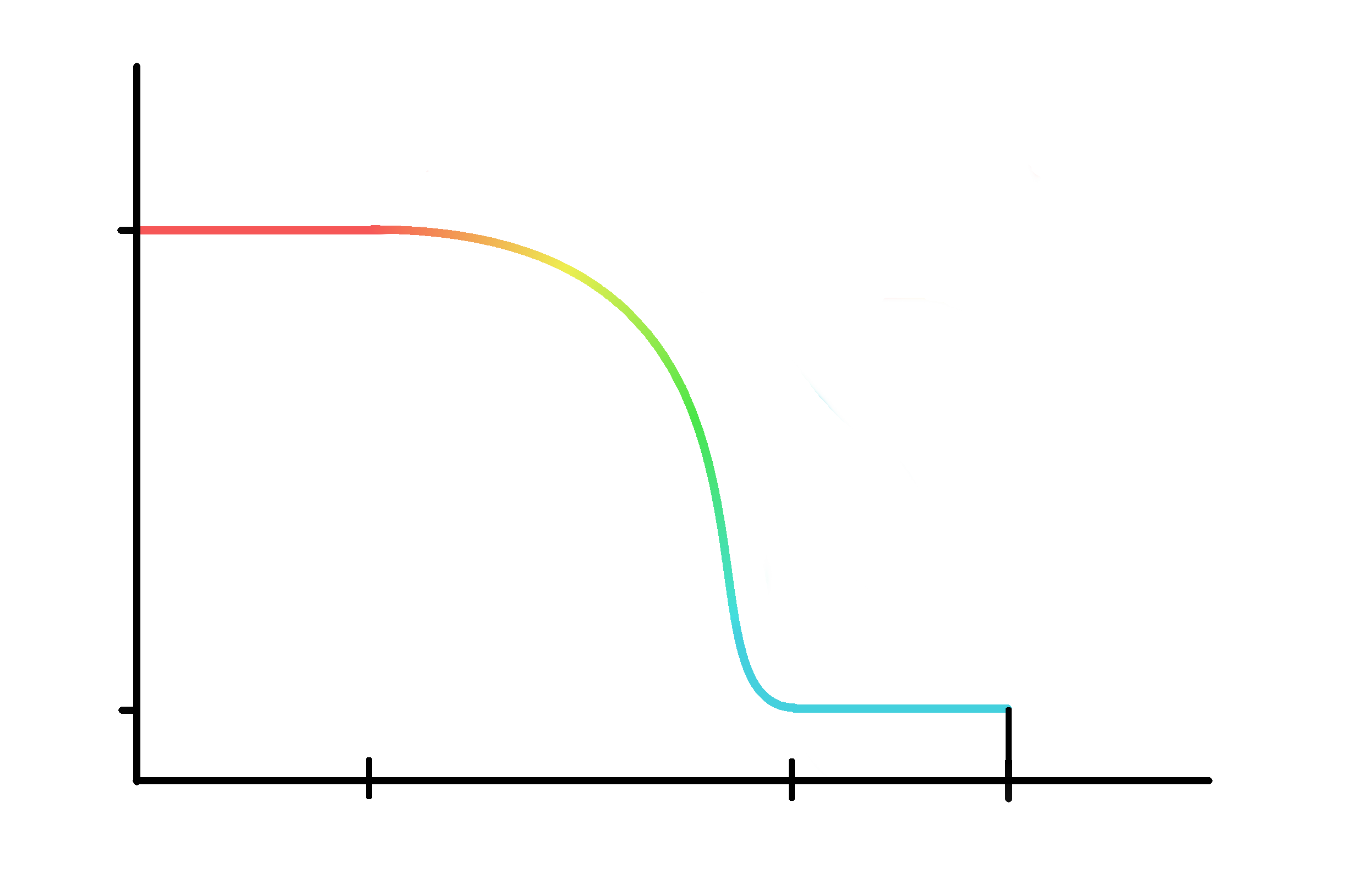}
\put (3,48) {$m$}
\put (3,13) {$m_1$}
\put (92,7) {$u$}
\put (40,52) {$M(u)$}
\put (15,3) {$II$}
\put (40,3) {$IV$}
\put (65,3) {$V$}
\put (80,3) {$VII$}
\end{overpic}
	\caption{Bondi-Sachs mass function along $\mathcal{J}^+$ for the toy model of evaporating black-to-white hole.}
	\label{mass-crossing-I}
\end{figure}
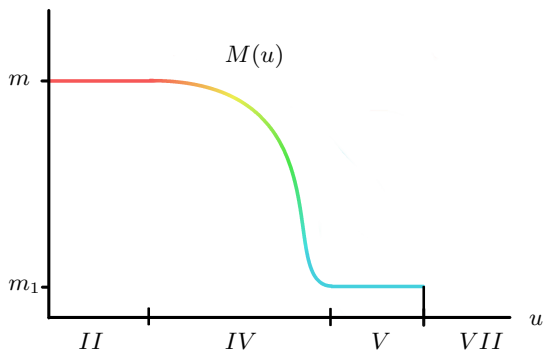

From the perspective of an observer lying on $\mathcal{J}^+$, the Bondi-Sachs mass evolves as depicted on the mass profile of figure \ref{mass-crossing-I}. It is positive and decreasing all along, going from $m$ to $0$. The white hole manifests itself through a sudden final release of positive energy corresponding to the emergence of the null bouncing shell. In region $III$ the inside Hawking quanta, which carry a negative energy, fades over the singularity, and never show up on the other side.

In \cite{Bianchi2014c, Bianchi2014a} it was shown that unitary evolution of an evaporating black hole implies a non-monotonic mass loss. To put it differently, a black hole must, at some point, radiate some amount of negative energy (the "last gasp"), which would be depicted on the mass profile as a momentary increase of the Bondi-Sachs mass. Intuitively, we can understand that the Hawking quanta, that fell inside the black hole, \textit{with negative energy}, are correlated with quanta outside, and should thus come out at some point, to recover unitarity on $\mathcal{J}^+$. The profile of Figure \ref{mass-crossing-I} does not fulfil the "last gasp" requirement. Indeed the flux of outgoing energy along $\mathcal{J}^+$ is
\begin{equation}
F(u) \propto - \frac{\dd M(u)}{\dd u},
\end{equation}
so that a momentary negative energy flux would mean a momentary increase of the Bondi-Sachs mass function.
However, preliminary discussions of De Lorenzo and Bianchi (personal communication), suggest that the last gasp theorem may fail in $4D$, in which case the mass profile of Figure \ref{mass-crossing-I} should not be discarded too easily.

Nevertheless, there is another reason why the previous model is not physically satisfying. For simplicity of the construction, we have assumed that the ingoing negative energy was fading along the singularity. Quantum gravity results suggests instead that it should cross the singularity. This calls for a refinement of our first toy model.

\vspace{\baselineskip}

{\bf Crossing model}

To do so, we consider that the Hawking quanta cross the singularity. It has been repeatedly noticed that there exists a natural prescription to extend geodesics beyond a singularity \cite{Synge1950, Peeters:1994jz, DAmbrosio2018}. Thus, modelling the crossing of the Hawking quanta through the singularity is the easy part of the refinement. It becomes more intricate afterwards. The negative energy flux is still ingoing, so it will fall upon the emerging bouncing shell. What comes next?

\begin{figure}[h]
\begin{overpic}[width =  0.5 \columnwidth]{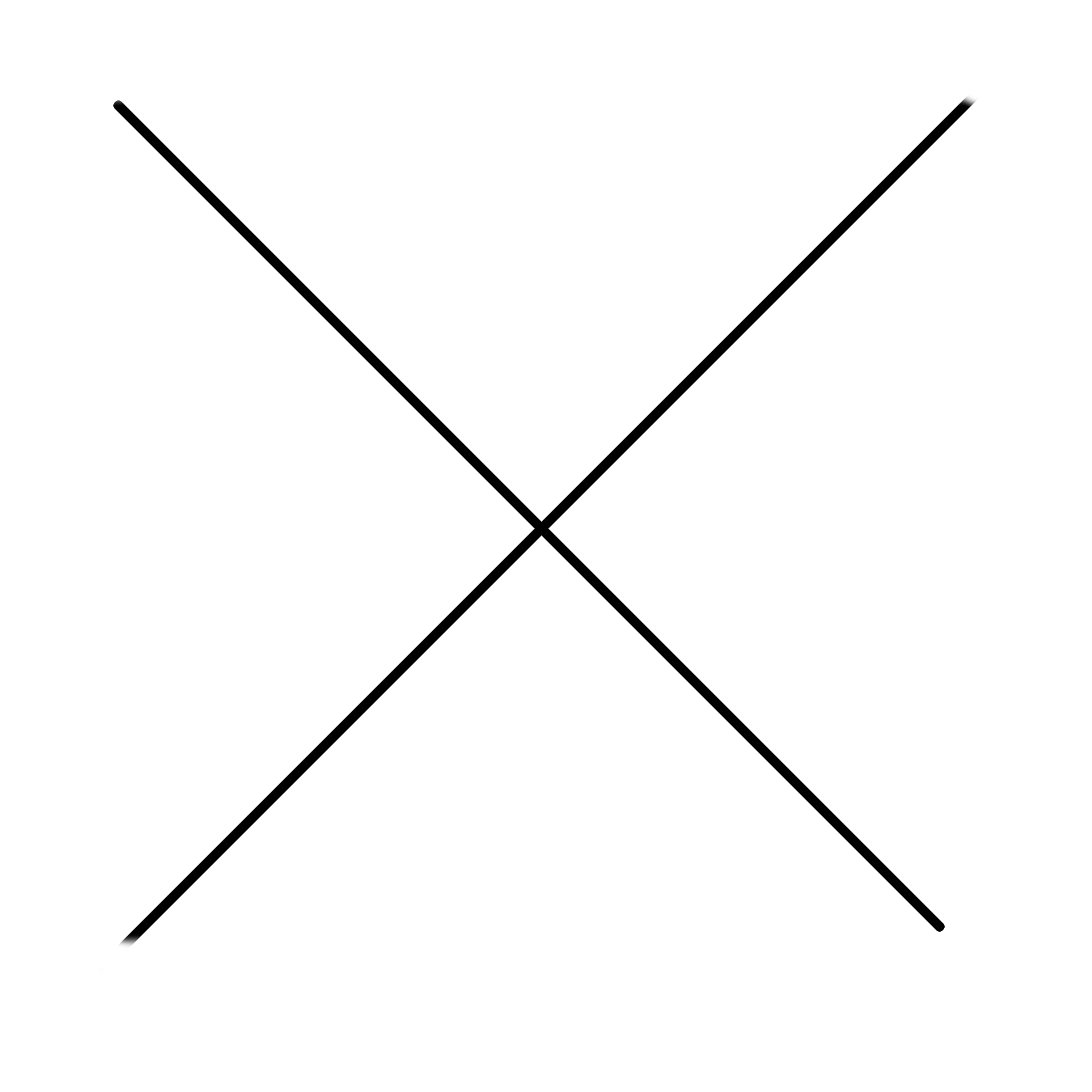}
\put (20,50) {$m_1$}
\put (80,50) {$m_2$}
\put (50,20) {$m_3$}
\put (50,80) {$m_4$}
\end{overpic}
	\caption{Four Schwarzschild patches can be glued consistently along null geodesics provided the masses satisfy equation \eqref{Hooft}.}
	\label{crossing}
\end{figure}
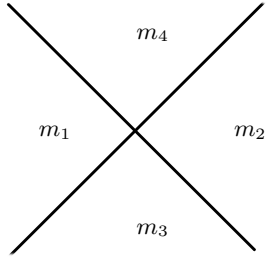

The crossing between two null shells has been studied by Dray and t'Hooft in \cite{Dray1985}. Their main result was that it was possible to glue four Schwatzschild patches along two null shells (see Figure \ref{crossing}), provided that the four masses satisfy the only condition
\begin{equation}\label{Hooft}
(r_0 - 2m_1)(r_0 - 2m_2) = (r_0 - 2m_3) (r_0 - 2m_4)
\end{equation}
where $r_0$ is the radius at the intersection.

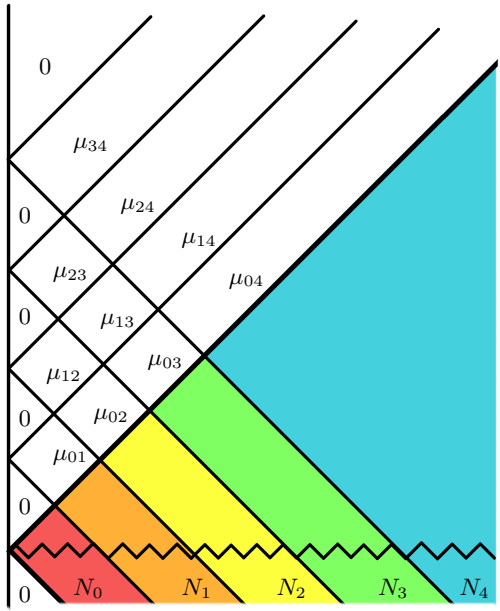
\begin{figure}[h]
\begin{overpic}[width =  1 \columnwidth]{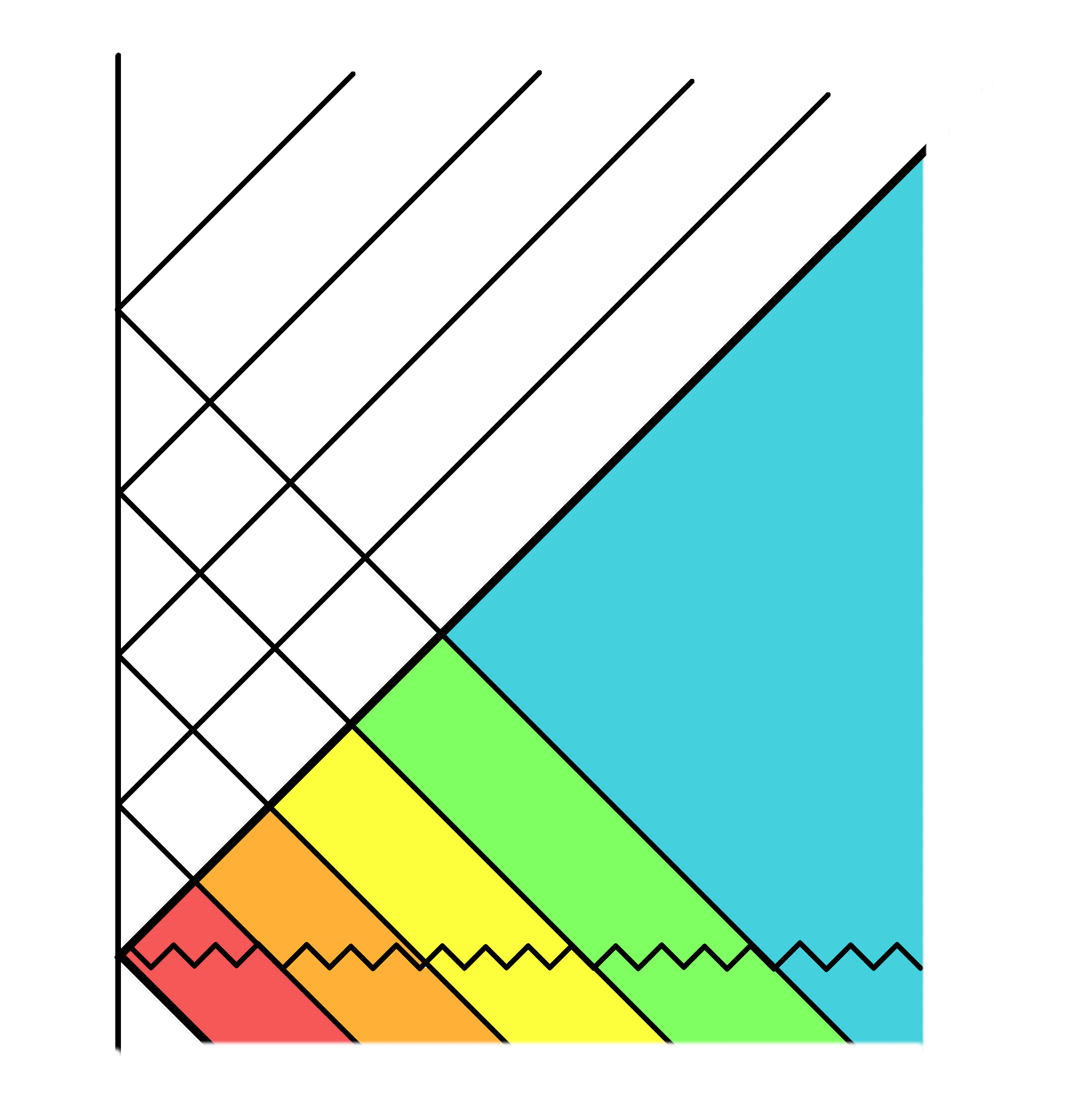}
\put (15,85) {$0$}
\put (12,63) {$0$}
\put (12,48) {$0$}
\put (12,33) {$0$}
\put (12,20) {$0$}
\put (12,7) {$0$}
\put (20,8) {$N_0$}
\put (36,8) {$N_1$}
\put (50,8) {$N_2$}
\put (65,8) {$N_3$}
\put (77,8) {$N_4$}
\put (17,28) {$\mu_{01}$}
\put (23,34) {$\mu_{02}$}
\put (31,42) {$\mu_{03}$}
\put (43,54) {$\mu_{04}$}
\put (16,40) {$\mu_{12}$}
\put (24,48) {$\mu_{13}$}
\put (36,60) {$\mu_{14}$}
\put (17,55) {$\mu_{23}$}
\put (27,65) {$\mu_{24}$}
\put (20,74) {$\mu_{34}$}
\end{overpic}
	\caption{Discrete model for the crossing between the ingoing Hawking quanta and the bouncing shell.}
	\label{mu-variations}
\end{figure}

To a first approximation, the ingoing flux, which was previously modelled continuously by a function $N(v)$, can be approached by a step function made of a number $n$ of slices of constant masses $N_i$. Then, the negative energy is carried by individual Hawking quanta which fall one at a time upon the bouncing shell. The situation is depicted on Figure \ref{mu-variations} for $n=5$.

In each box, the metric is Schwarzschild with a constant mass $\mu_{ij}$, which is determined by equation \eqref{Hooft} as a function of the three masses in the adjacent boxes below and the value of the radius where the four regions touch. Now, suppose the radius in the region above the bouncing shell is increasing (resp. decreasing) along outgoing (resp. ingoing) null geodesics. We prove the following theorem:

\vspace{\baselineskip}
{\bf Theorem 1.} {\it Under the above assumptions, $\mu_{ij}$ is a decreasing function of $i$ and an increasing function of $j$. This implies notably that for all $i,j$ 
\begin{equation}
0 < \mu_{ij} < m_1.
\end{equation}}

\vspace{\baselineskip}
{\sc Proof.}
Let us first study $\mu_{0j}$ for varying $j$. Denote $r_j$ the radius at the intersection point at the bottom of the box of mass $\mu_{0j}$. From equation \eqref{Hooft}, we deduce
\begin{equation}
(\mu_{0,j} - \mu_{0,j-1}) (r_j - 2N_j) = (N_j - N_{j-1}) (r_j - 2 \mu_{0,j})
\end{equation}
Since $N_j$ is decreasing, we have $N_j < N_{j-1}$. Then, since the radius $r_j$ is assumed to be increasing, we have $r_j \leq 2 m_1 \leq 2N_j$, so that 
\begin{equation}\label{ineq-mu-r}
\mu_{0,j} > \mu_{0,j-1} \Leftrightarrow r_j > 2 \mu_{0,j}.
\end{equation}
which can be restated saying that for each $j$, one, and only one of the two following must hold:
\begin{equation}
\begin{array}{l}
\mu_{0,j-1} < \mu_{0,j}  < \frac{r_j}{2}  \\
\mu_{0,j-1} > \mu_{0,j}  > \frac{r_j}{2}.
\end{array}
\end{equation}
Initially, we have $r_0 = 0$. Since $r_1 >0$ we deduce
\begin{equation}
 0 < \mu_{01} < \frac{ r_1}{2}.
\end{equation}
Then, using that $r_{j+1} > r_j$, we show by induction that for any $j$
\begin{equation}
\mu_{0,j-1} < \mu_{0,j}  < \frac{r_j}{2} .
\end{equation}
Thus $\mu_{0j}$ is increasing with $j$ and satisfies
\begin{equation}
0 < \mu_{0j}  < m_1 .
\end{equation}

A similar induction shows that $\mu_{1j}$ is also an increasing function of $j$, satisfying.
\begin{equation}
0 < \mu_{1j} < m_1 .
\end{equation}

Then, under the assumption of decreasing $r$ along ingoing null geodesics, an induction over $i$ shows that for any $j$, $\mu_{ij}$ is a decreasing function of $i$.
$\square$

\vspace{\baselineskip}

The previous discrete model gives a fair description of what can happen when a series of Hawking quanta successively cross the bouncing shell. In the continuum limit, when $n \to \infty$, the resulting metric takes the form
\begin{equation}
\dd s^2 = - \left(1-\frac{2 \mu(u,v)}{r}\right) du dv + r^2 d\Omega^2
\end{equation}
characterised by two functions, namely the radius $r(u,v)$ and the mass $\mu(u,v)$. We cannot give explicitely the change of coordinates from $(u,v)$ to $(U,V)$ but we assume that $v(V)$ and $u(U)$ are increasing. Then, theorem 1 shows that
\begin{equation}
\frac{\partial \mu}{\partial u} < 0  \qquad \text{and} \qquad \frac{\partial \mu}{\partial v} > 0 .
\end{equation}
As a corollary we have
\begin{equation}
0 < \mu(u,v) < m_1.
\end{equation}
We have no explicit expression neither for the radius $r(u,v)$ nor for the mass $\mu(u,v)$, for it would require integrate too difficult equations. However, it is clear for the construction of the discrete setting above that such functions exist.

\begin{figure}[h]
\begin{overpic}[width =  0.9 \columnwidth]{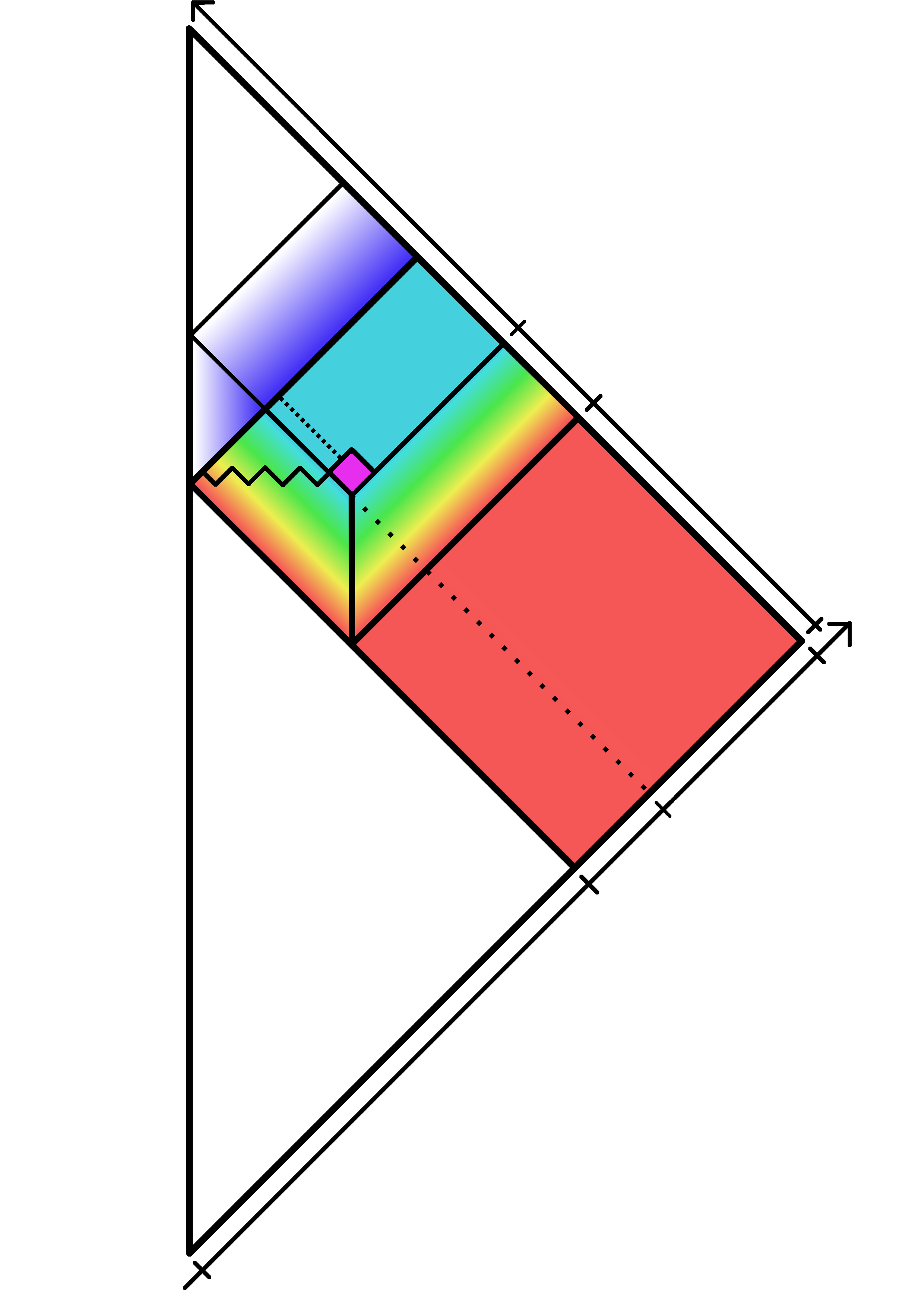}
\put (25,35) {$I$}
\put (45,50) {$II$}
\put (20,58) {$III$}
\put (33,65) {$IV$}
\put (30,70) {$V$}
\put (23,66) {\tiny $a$}
\put (18,65) {\tiny $VIb$}
\put (15,68) {\tiny $VII$}
\put (20,77) {$VIII$}
\put (17,85) {$IX$}
\put (68,50) {$V$}
\put (63,45) {$\frac{\pi}{2}$}
\put (52,35) {$V_0-\varepsilon $}
\put (48,28) {$\frac{\pi}{4} $}
\put (18,0) {$\frac{\pi}{2} - 2 V_0 $}
\put (10,100) {$U$}
\put (34,84) {$ 2V_0 - \frac{3\pi}{4}$}
\put (40,76) {$ V_0 -\varepsilon- \frac{\pi}{2}$}
\put (48,70) {$- \frac{\pi}{4}$}
\put (61,55) {$- \frac{\pi}{2}$}
\end{overpic}
	\caption{Penrose diagram of an evaporating black-to-white hole with ingoing energy flux that crosses first the singularity and then the bouncing shell. The dashed boundary $V/VIa$ represents the apparent horizon of the white hole, characterised by $r= 2m_1$.}
	\label{crossing-II}
\end{figure}

To sum-up, the resulting space-time is depicted on Figure \ref{crossing-II}, with the metric given by
\begin{align}
(V) &
\left[
\begin{array}{l}
\dd s^2 = - \left(1-\frac{2m_1}{r}\right) du dv + r^2 d\Omega^2\\
r = 2m_1 \left( 1+ W\left( e^{\frac{v-u}{4m_1} -1} \right) \right) 
\end{array}
\right. \\
(VIa) &
\left[
\begin{array}{l}
\dd s^2 = \left(1-\frac{2m_1}{r}\right) du dv + r^2 d\Omega^2\\
r = 2m_1 \left( 1+ W\left( - e^{- \frac{v+u}{4m_1} -1} \right) \right) 
\end{array}
\right. \\
(VIb) &
\left[
\begin{array}{l}
\dd s^2 = - \left(1-\frac{2\tilde{N}(v)}{r}\right) dv^2 + 2dv dr + r^2 d\Omega^2 
\end{array}
\right. \\
(VII) &
\left[
\begin{array}{l}
\dd s^2 = - \left(1-\frac{2 \mu(u,v)}{r}\right) du dv + r^2 d\Omega^2
\end{array}
\right. \\
(VIII) &
\left[
\begin{array}{l}
\dd s^2 =  - \left(1-\frac{2P(u)}{r}\right) du^2 - 2du dr + r^2 d\Omega^2 
\end{array}
\right. \\
(IX) &
\left[
\begin{array}{l}
\dd s^2 = - du dv + r^2 d\Omega^2 \\
r= \frac{1}{2} \left( v - u \right)
\end{array}
\right. 
\end{align}
In regions $I-IV$ the metric is the same as the model of Hiscock (see equations (\ref{Hiscock metric I}-\ref{Hiscock metric IV}) and (\ref{HiscockII-Ia}-\ref{HiscockII-IV})). The mass function $\tilde{N}(v)$ that appears in the metric of region $VIb$ is chosen to match the mass function $N(v)$ along the boundary $III/VIb$. Similarly, the mass function $P(u)$ of region $VIII$ is chosen to match $\mu(u,v)$ along the boundary $VII/VIII$. The map between the coordinates $(u,v)$ and $(U,V)$ cannot be given explicitly. 

A word shall be added concerning the size of the central diamond region. The future endpoint of the apparent horizon of the black hole has a radius $r=2m_1$, which characterises the typical size of the diamond. The mathematical construction of the model requires that $0<m_1<m$. However, physically, $m_1$ is believed to be small. How small? Well, remember that in quantum gravity the singularity is expected to be actually a "thick" singularity, i.e. a Planck star whose radius is given by $r \sim N(v)^{1/3}$. The power $1/3$ is obtained from the condition that the curvature should become Planckian. So a Planck star can actually be quite big. Now, along the apparent horizon, the radius is given by $r \sim 2 N(v)$. Then, evaporation can last at most until the "thick singularity" and the apparent horizon meet, i.e. when $m_1^{1/3} \sim 2 m_1$. This condition means the mass $m_1$ should be Planckian. Without surprise, a Planckian $m_1$ thus marks a lower bound for our model. In this case, the size of the diamond itself is Planckian, so it is really just one quantum of space. 

\begin{figure}[h]
\begin{overpic}[width = .9 \columnwidth]{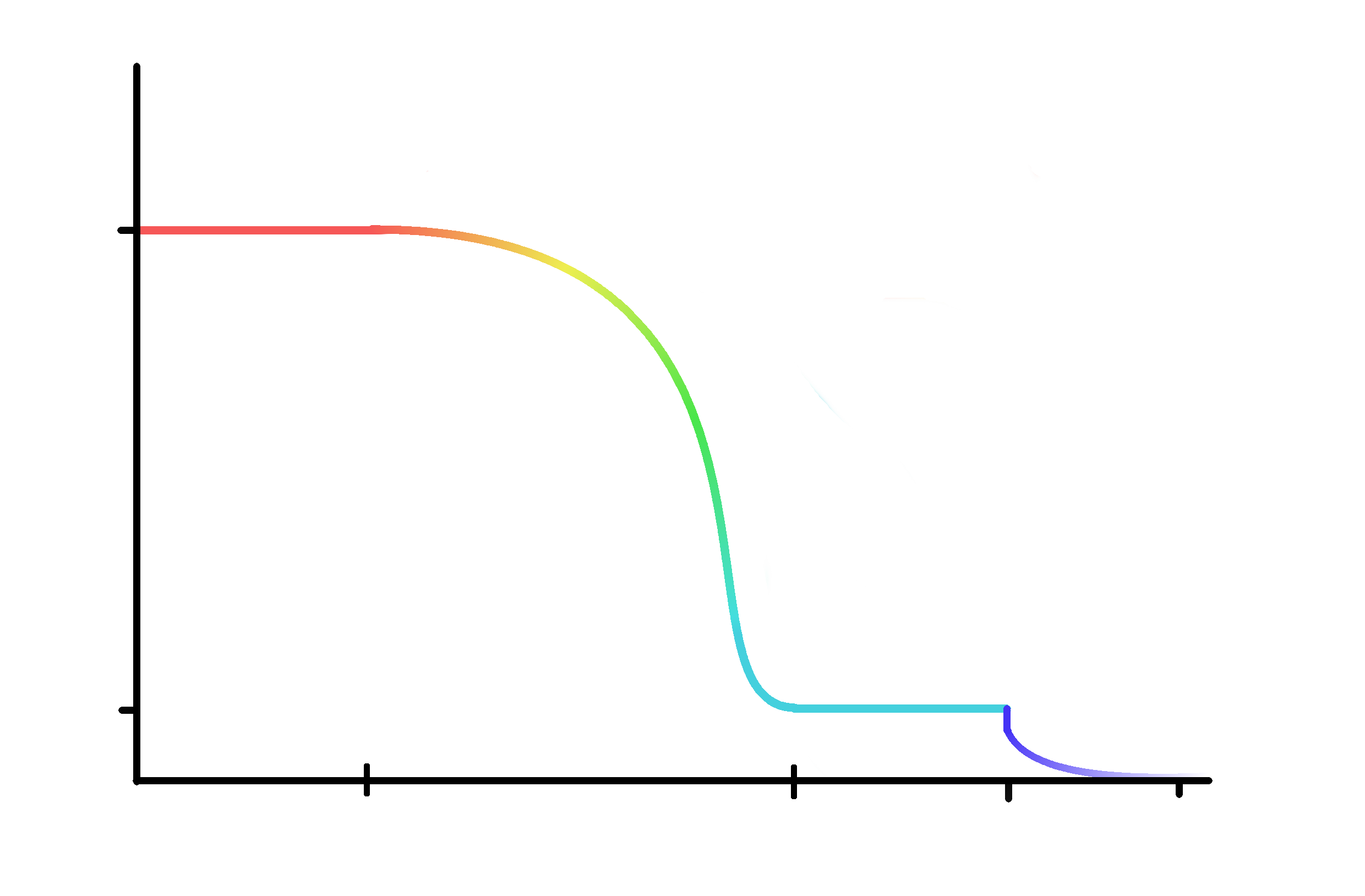}
\put (3,48) {$m$}
\put (3,13) {$m_1$}
\put (92,7) {$u$}
\put (15,3) {$II$}
\put (40,3) {$IV$}
\put (65,3) {$V$}
\put (76,3) {$VIII$}
\put (89,3) {$IX$}
\end{overpic}
	\caption{Bondi-Sachs mass function along $\mathcal{J}^+$ for a refined model of evaporating black-to-white hole.}
	\label{mass-ingoing}
\end{figure}

The resulting mass profile along $\mathcal{J}^+$ is shown on Figure \ref{mass-ingoing}.
Instead of a sharp release of energy when the shell bounces out, as in the previous toy model (see Figure \ref{mass-crossing-I}), the Bondi-Sachs mass slowly decreases to zero. It can be interpreted as the emergence of the Hawking quanta that finally reach $\mathcal{J}^+$. It should be noticed however that they carry positive energy (since the Bondi-Sachs mass is decreasing all along), whereas they were known to carry negative energy after they formed at the apparent horizon. This change of sign is due to the exchange of energy that occurs when the quanta cross the bouncing shell: positive energy from the shell is transfered to the quanta. The final long-dying tail on the mass profile enables energy (and information) to be slowly released. 

\section{Model of evaporating black-to-white hole (II)}\label{section model II}

{\bf \textit{Outgoing} inside radiation}

As can be seen from equations \eqref{Hawking flux}, Hawking flux is outgoing even inside the black hole. In other words, Hawking quanta are well falling towards the singularity, but they are \textit{out}-falling, i.e. falling along outgoing null geodesics. This has lead some people to doubt the credibility of the previous Hiscock model, where the correction inside the hole only corresponds to an \textit{in}-falling negative energy flux. Nevertheless, this objection is not correct because the iterative approach to the semi-classical Einstein equations requires to consider the full $\langle in | T_{\mu \nu} | in \rangle$, including both the Hawking flux contribution $\langle in | :T_{\mu \nu} :| in \rangle$ and the vacuum polarization part $\langle B | T_{\mu \nu} | B \rangle$. The formulae are given by equations (\ref{T_uu}-\ref{T_uv}) and we see that all of the components play a role. 

We justified Hiscock model earlier by looking at the direction of the flux along the horizon and along $\mathcal{J}^+$. We noticed that along the horizon, the only non-vanishing component is $\expval{T_{vv}}$, which corresponds to an ingoing flux. However it is true that, as we move away from the horizon, towards the singularity, the components $\expval{T_{uu}}$ and $\expval{T_{uv}}$ come into play. In fact, on the singularity itself, when $r \to 0$, all the components of $\expval{T_{\mu \nu}}$ diverge, but with the same behaviour: 
\begin{equation}\label{Tmunu singularity}
\begin{split}
\expval{T_{uu}} \sim  - \frac{\hbar}{24 \pi} \frac{m}{r^3} \\
\expval{T_{vv}} \sim - \frac{\hbar}{24 \pi} \frac{m}{r^3} \\
\expval{T_{uv}} \sim - \frac{\hbar}{24 \pi} \frac{m}{r^3}.
\end{split}
\end{equation}

\begin{figure}[h]
\begin{overpic}[width =  0.9 \columnwidth]{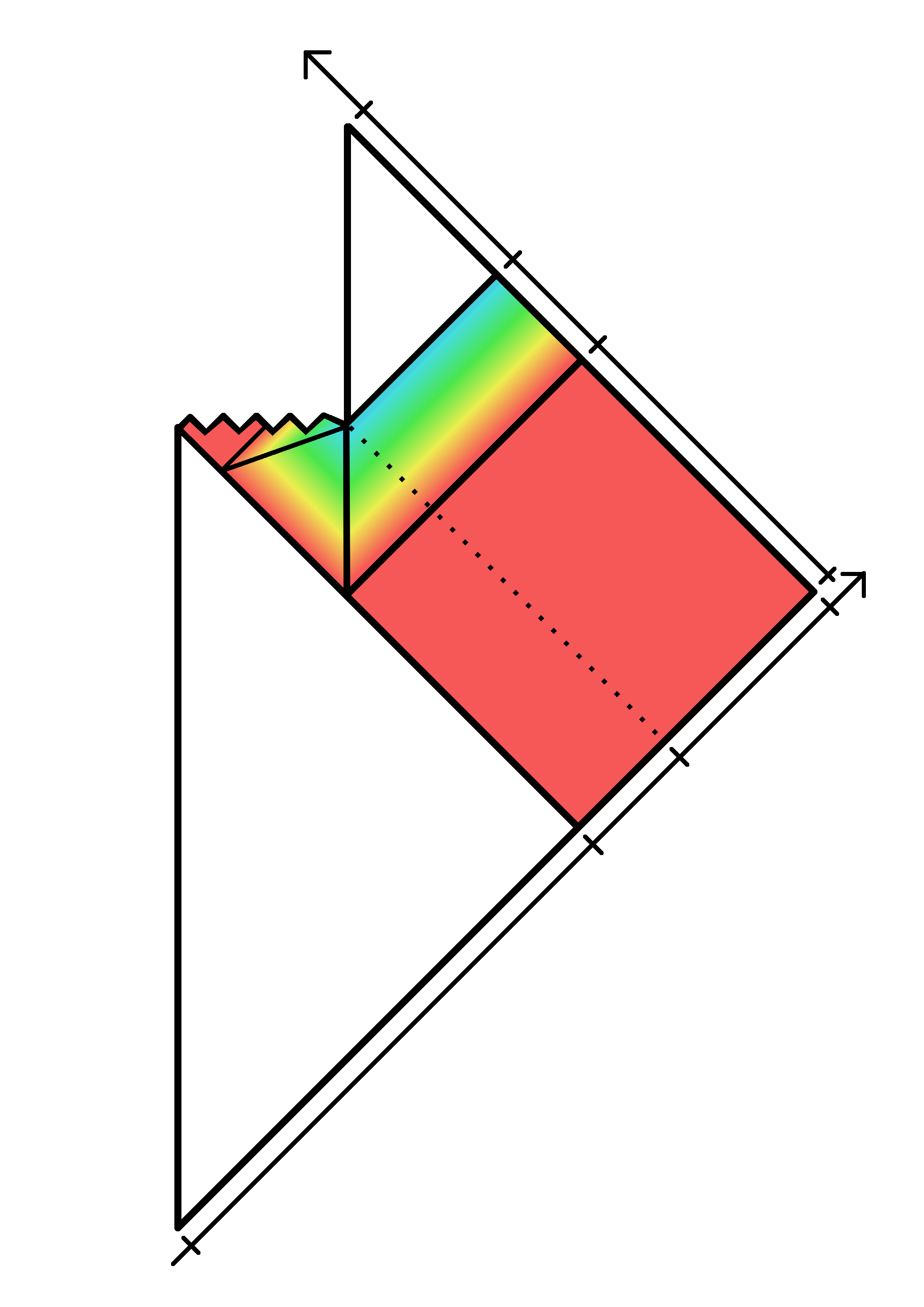}
\put (25,35) {$I$}
\put (45,55) {$II$}
\put (20,62) {$III$}
\put (33,67) {$IV$}
\put (30,78) {$V$}
\put (21,71) {$VI$}
\put(22,70){\vector(0,-1){3.8}}
\put (13,71) {$VII$}
\put(17,70){\vector(0,-1){4}}
\put (68,55) {$V$}
\put (65,50) {$\frac{\pi}{2}$}
\put (55,40) {$V_0 $}
\put (48,32) {$\frac{\pi}{4} $}
\put (18,3) {$\frac{\pi}{2} - 2 V_0 $}
\put (18,95) {$U$}
\put (30,93) {$0$}
\put (40,82) {$ V_0 - \frac{\pi}{2}$}
\put (48,75) {$- \frac{\pi}{4}$}
\put (61,60) {$- \frac{\pi}{2}$}
\end{overpic}
	\caption{Penrose diagram of an evaporating black hole with inside \textit{outgoing} flux.}
	\label{alterHiscock}
\end{figure}
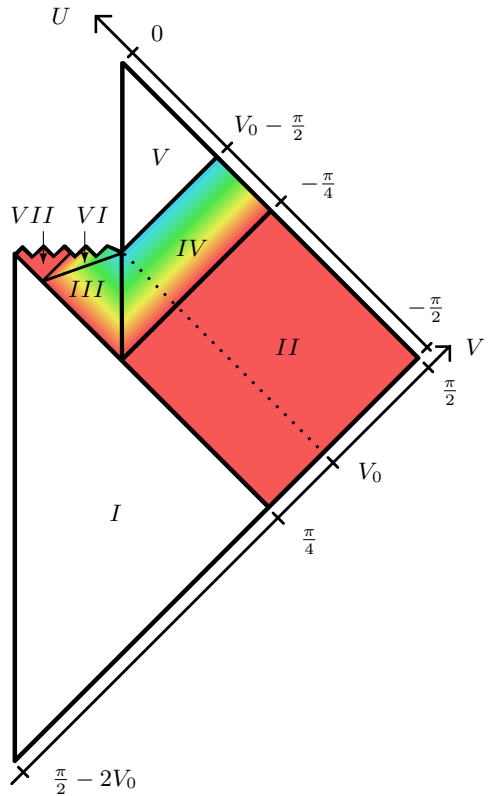

One of the goals of this article is to investigate the fate of the negative energy after it has reached the singularity, in a black-to-white scenario. To that aim, the direction of the energy when it reaches the singularity matters. We have thus considered equally important to study the case of an outgoing energy flux inside the hole. This has motivated the design of another model of evaporating black hole. It is a slight modification of Hiscock model inside the hole. The idea is simple: after the ingoing flux of particles has been created along the apparent horizon, as in the Hiscock model, they are scattered by the gravitational field, and change direction. This scattering is sketched by introducing a space-like surface inside the hole (boundary $III/VI$) along which particles are deviated. The model is represented as a Penrose diagram on Figure \ref{alterHiscock}.

The metric is given in 7 patches. In regions $I-V$, the metric is the same as Hiscock model, given by equations (\ref{Hiscock metric I}-\ref{Hiscock metric V}). The space-like boundary $III/VI$ is chosen arbitrarily. In regions $VII$ and $VIII$ the metric is given by
\begin{align}
\label{outgoing metric VI} (VI) &
\left[
\begin{array}{l}
\dd s^2 = - \left(1-\frac{2Q(u)}{r}\right) du^2 - 2du dr + r^2 d\Omega^2 
\end{array}
\right. \\
\label{outgoing metric VII} (VII) &
\left[
\begin{array}{l}
\dd s^2 =  \left(1-\frac{2m}{r}\right) du dv + r^2 d\Omega^2 \\
r = 2m \left( 1+ W\left(- e^{\frac{v+u}{4m} -1} \right) \right) 
\end{array}
\right. 
\end{align}
The mass function $Q(u)$ is chosen so that it matches the mass function $N(v)$ along the boundary $III/VI$. The metric is written in terms of coordinates $(u,v)$ or $(u,r)$, which are related to the coordinates $(U,V)$ of the Penrose diagram by the formulae in the appendix \ref{detail inside outgoing}.

\begin{figure}[h]
\begin{overpic}[width =  0.9 \columnwidth]{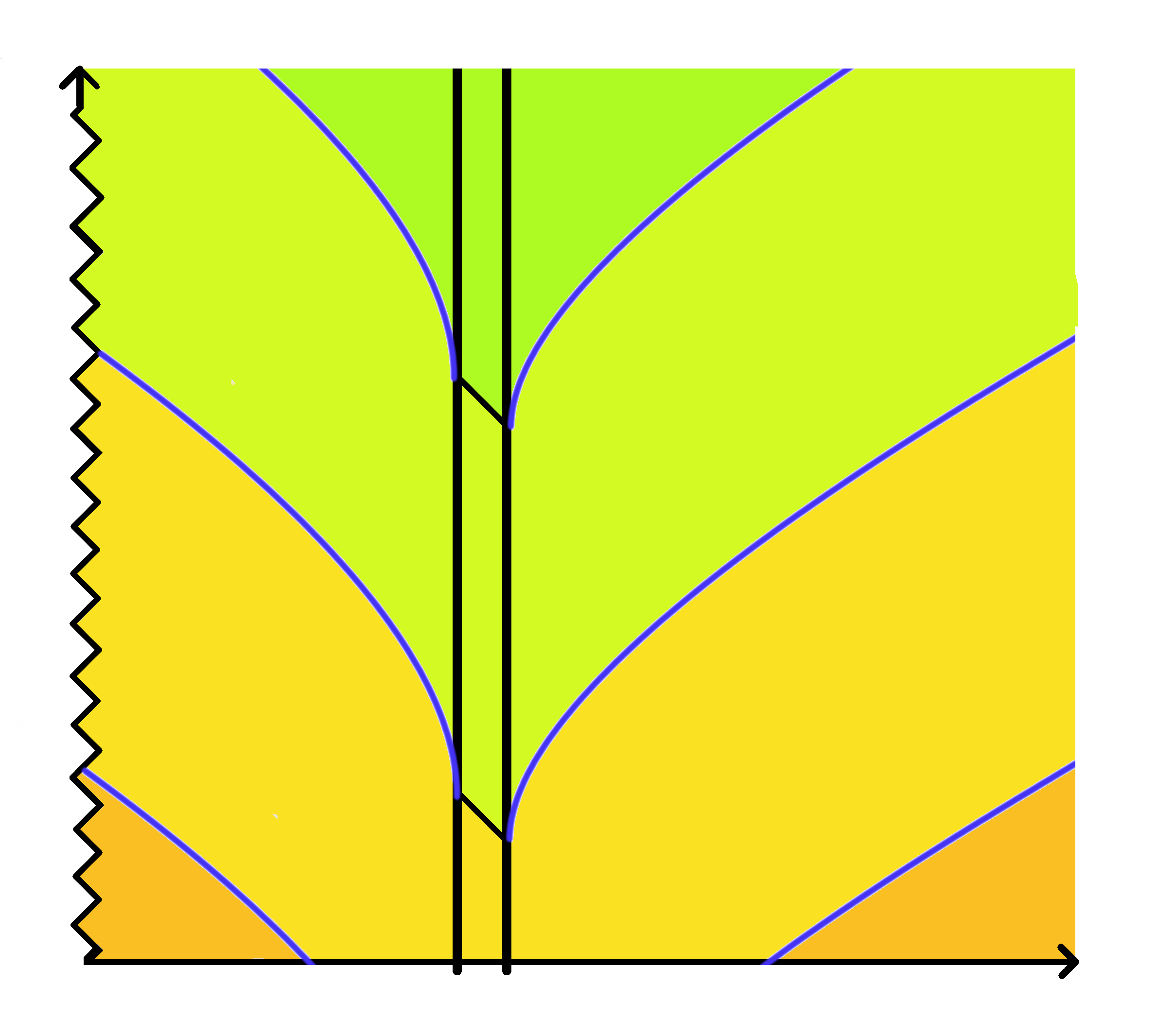}
\put (20,0) {$VI$}
\put (39,0) {$III$}
\put (60,0) {$IV$}
\put (0,80) {$\tilde{t}$}
\put (95,5) {$r$}
\end{overpic}
	\caption{The region surrounding the horizon in Eddington time coordinate. Three pairs of Hawking quanta are represented by blue lines.}
	\label{eddington}
\end{figure}
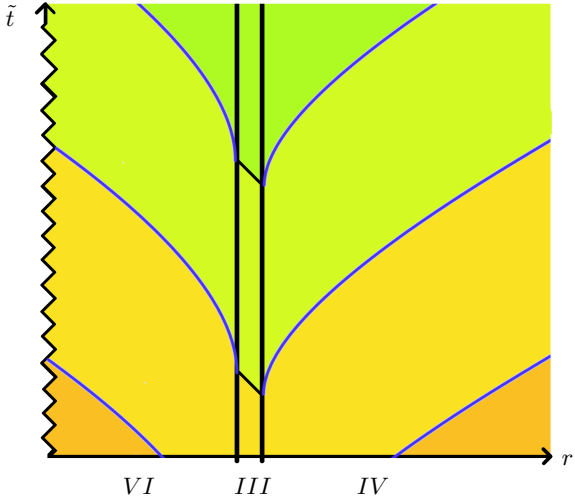

Above, we have introduced the outgoing flux as a consequence of the scattering of the ingoing flux. Another, maybe simpler, physical intuition can be given for the outgoing flux provided a better system of coordinates is used to represent the region surrounding the horizon. Indeed, due to the distortion of distances, the Penrose diagram does not properly depict the fact that the space-like boundary $III/VI$ and the time-like apparent horizon $III/IV$ may actually be very close. Using Eddington time coordinate, \begin{equation}
\tilde{t} = t + 2m \log \left| \frac{r}{2m} - 1 \right|  \quad \text{with} \ t = \frac{u+v}{2},
\end{equation}
a small region around the horizon, thin in $V$, but large enough in $U$ to include the two boundaries, looks like Figure \ref{eddington}.
Regions $IV$ and $VI$ surround a very small region $III$. Pairs of Hawking quanta are created alongside the null event horizon. Both quanta, inside and outside the hole, are outgoing, i.e. following the same side of the light cone (remember that the light cones are tilted in the Eddington time representation). However sketchy this description may be, we see that the modified model proposed in this section, with outoing inside radiation, can be related to the usual intuitive idea of pairs of particles created along the event horizon.

\vspace{1 \baselineskip}

{\bf Evaporating black-to-white hole (II)}

\begin{figure}[h]
\begin{overpic}[width =  0.9 \columnwidth]{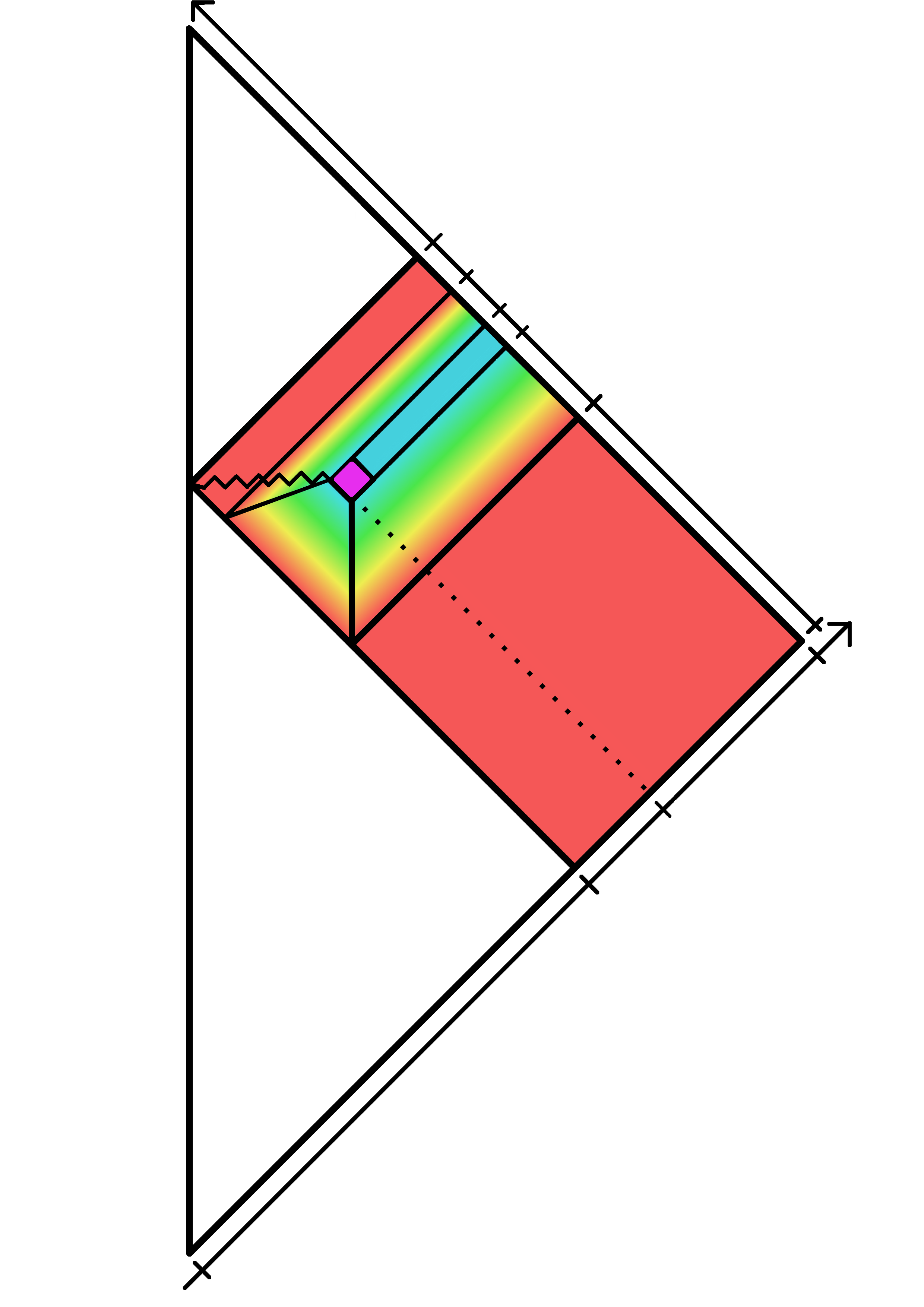}
\put (55,90) {$\mathcal{J}^+$}
\put(10,62){\vector(1,0){7}}
\put (5,61) {$VI$}
\put(10,52){\vector(1,1){10}}
\put (4,51) {$VII$}
\put (25,35) {$I$}
\put (45,50) {$II$}
\put (20,58) {$III$}
\put (33,63) {$IV$}
\put (30,66) {\scriptsize $V$}
\put (27,70) {\tiny $VIII$}
\put (20,67) {\tiny $IX$}
\put (20,80) {$X$}
\put (68,50) {$V$}
\put (63,45) {$\frac{\pi}{2}$}
\put (52,35) {$V_0-\varepsilon $}
\put (48,28) {$\frac{\pi}{4} $}
\put (18,0) {$\frac{\pi}{2} - 2 V_0 $}
\put (10,100) {$U$}
\put (34,84) {$ 2V_0 - \frac{3\pi}{4}$}
\put (41,74) {\scriptsize $ V_0 -\varepsilon- \frac{\pi}{2}$}
\put (39,77) {\scriptsize $ V_0 + \varepsilon- \frac{\pi}{2}$}
\put (48,70) {$- \frac{\pi}{4}$}
\put (61,55) {$- \frac{\pi}{2}$}
\end{overpic}
	\caption{}
	\label{B2Wnew}
\end{figure}

The new model proposed for an evaporating black hole extends naturally to the black-to-white hole scenario. The inside energy flux cross the singularity, and goes ahead towards $\mathcal{J}^+$. The corresponding Penrose diagram is drawn on Figure \ref{B2Wnew}.
The metric in regions $I-IV$ is given by the equations (\ref{Hiscock metric I}-\ref{Hiscock metric IV}). In regions $VI$ and $VII$ it is given by equations \eqref{outgoing metric VI} and \eqref{outgoing metric VII}. Elsewhere, the metric is given by
\begin{align}
(V) &
\left[
\begin{array}{l}
\dd s^2 = - \left(1-\frac{2m_1}{r}\right) du dv + r^2 d\Omega^2\\
r = 2m_1 \left( 1+ W\left( e^{\frac{v-u}{4m_1} -1} \right) \right) 
\end{array}
\right. \\
(VIII) &
\left[
\begin{array}{l}
\dd s^2 =  - \left(1-\frac{2R(u)}{r}\right) du^2 - 2du dr + r^2 d\Omega^2 
\end{array}
\right. \\
(IX) &
\left[
\begin{array}{l}
\dd s^2 = - \left(1-\frac{2m}{r}\right) du dv + r^2 d\Omega^2\\
r = 2m \left( 1+ W\left( e^{\frac{v-u}{4m} -1} \right) \right) 
\end{array}
\right. \\
(X) &
\left[
\begin{array}{l}
\dd s^2 = - du dv + r^2 d\Omega^2 \\
r= \frac{1}{2} \left( v - u \right)
\end{array}
\right. 
\end{align}
The mass function $R(u)$ is such that it matches with that of region $VII$ along the singularity.
The metric inside the purple central diamond has been already discussed in section \ref{section model I}. We do not give explicitely the map between the coordinates $(u,v)$ and $(U,V)$, but there is not doubt that the construction is consistent, and that the gluing can be performed along all boundaries.

\begin{figure}[h]
\begin{overpic}[width = .9 \columnwidth]{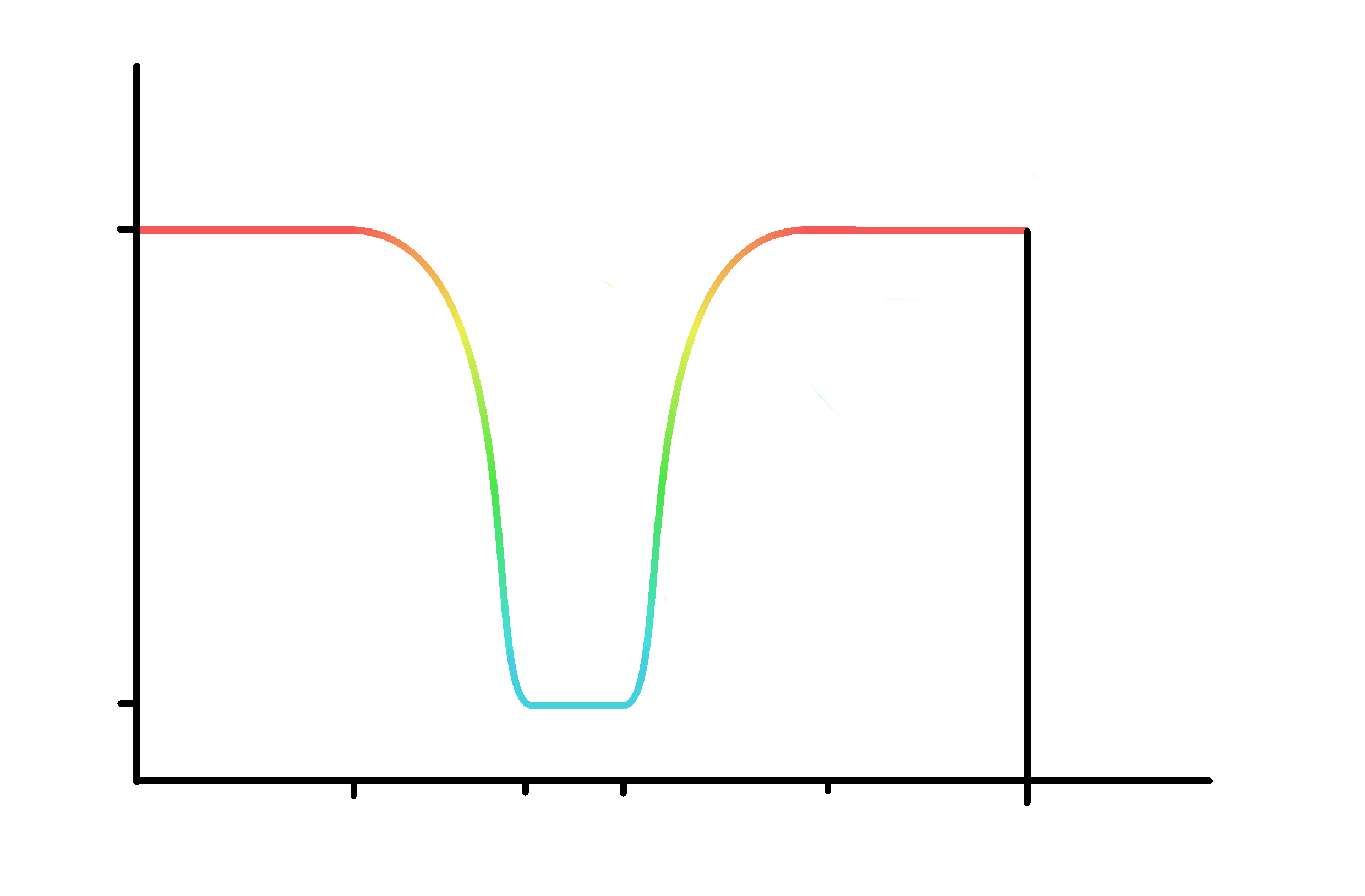}
\put (3,48) {$m$}
\put (3,13) {$m_1$}
\put (92,7) {$u$}
\put (15,3) {$II$}
\put (30,3) {$IV$}
\put (40,3) {$V$}
\put (50,3) {$VIII$}
\put (65,3) {$IX$}
\put (80,3) {$IX$}
\end{overpic}
	\caption{Bondi-Sachs mass function along $\mathcal{J}^+$ for a model of evaporating black-to-white hole with outgoing inside radiation.}
	\label{mass-outgoing}
\end{figure}
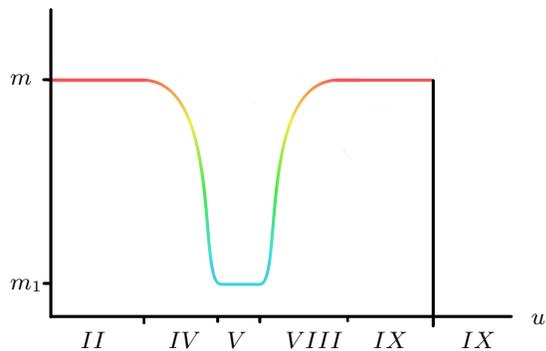

In this scenario, the Bondi-Sachs mass is shown on Figure \ref{mass-outgoing}.
Contrary to the previous scenarios, the mass function is not monotonic: after the black hole has shrinked from $m$ to $m_1$, the transition to a white hole occurs, and then the mass increases again from $m_1$ to $m$. All these outgoing quanta carry a negative energy, so that the energy conditions are strongly violated in this case. This feature makes the scenario consistent with the expectation of a "last gasp" \cite{Bianchi2014a, Bianchi2014c}, but the violation is clearly too strong to be physically acceptable. The information loss paradox is obviously solved since the inside quanta, which are correlated to those emitted outside the hole, finally reach $\mathcal{J}^+$. The model proposed by James M. Bardeen in \cite{Bardeen2018}, is quite similar to this second model presented here.

In this over-simplified model, all inside quanta are outgoing while it is known from equations \eqref{Tmunu singularity} that only part of them reaches the singularity with this direction. A fully satisfying model would then lie in-between the ingoing model of section \ref{section model I} and the outgoing one of section \ref{section model II}. As a result, the mass profile itself should lie somewhere between Figure \ref{mass-ingoing} and Figure \ref{mass-outgoing}.

\section{Conclusion}

In this article, we have constructed and discussed several effective models that describe an evaporating black-to-white hole. Based on a first construction by Hiscock, we have emphasised the double contribution from vacuum polarisation and Hawking quanta to the expectation value of the energy-momentum tensor that enters the semi-classical Einstein equations. This justifies that we should consider both models where the inside radiation is ingoing and outgoing. Then, we have shown how an evaporating black hole can be naturally extended to a white hole future, as quantum gravity suggests. Whereas the black-to-white hole model proposed in \cite{Rovelli2018a} was flawed by the well-known instability of white holes, it is a nice feature of the evaporating model to cure it, as already noticed in \cite{Bianchi2018}. The consistent mathematical models finally obtained display two main different profiles for the Bondi-Sachs mass along $\mathcal{I}^+$, but it is believed that the actual phenomenology should lie in-between the two.

If it exists, the black-to-white hole transition is thought to be a quantum tunnelling phenomenon. It is thus expected of any theory of quantum gravity to provide tools to compute the quantum amplitude of the transition. Loop Quantum Gravity offers such tools, relying over the definition of a boundary surrounding the region where quantum effects are expected to be dominant. In our models, this region is a central diamond, and we have shown how its size could be reduced to Planckian scale. We let to future works the task of effectively computing the transition amplitude. Such a computation would ultimately confirm or not previous estimations of the probabilty of transition and the lifetime of black holes.

\section*{Acknowledgements}

We thank Tommaso De Lorenzo, Alejandro Perez and James M. Bardeen for useful exchanges. We acknowledge the OCEVU Labex (ANR-11-LABX-0060) and the A*MIDEX project (ANR-11-IDEX-0001-02) funded by the "Investissements d'Avenir" French government program managed by the ANR.

\appendix

\section{Details of Hiscock model}\label{details Hiscock}

\begin{figure}[h]
\begin{overpic}[width =  0.9 \columnwidth]{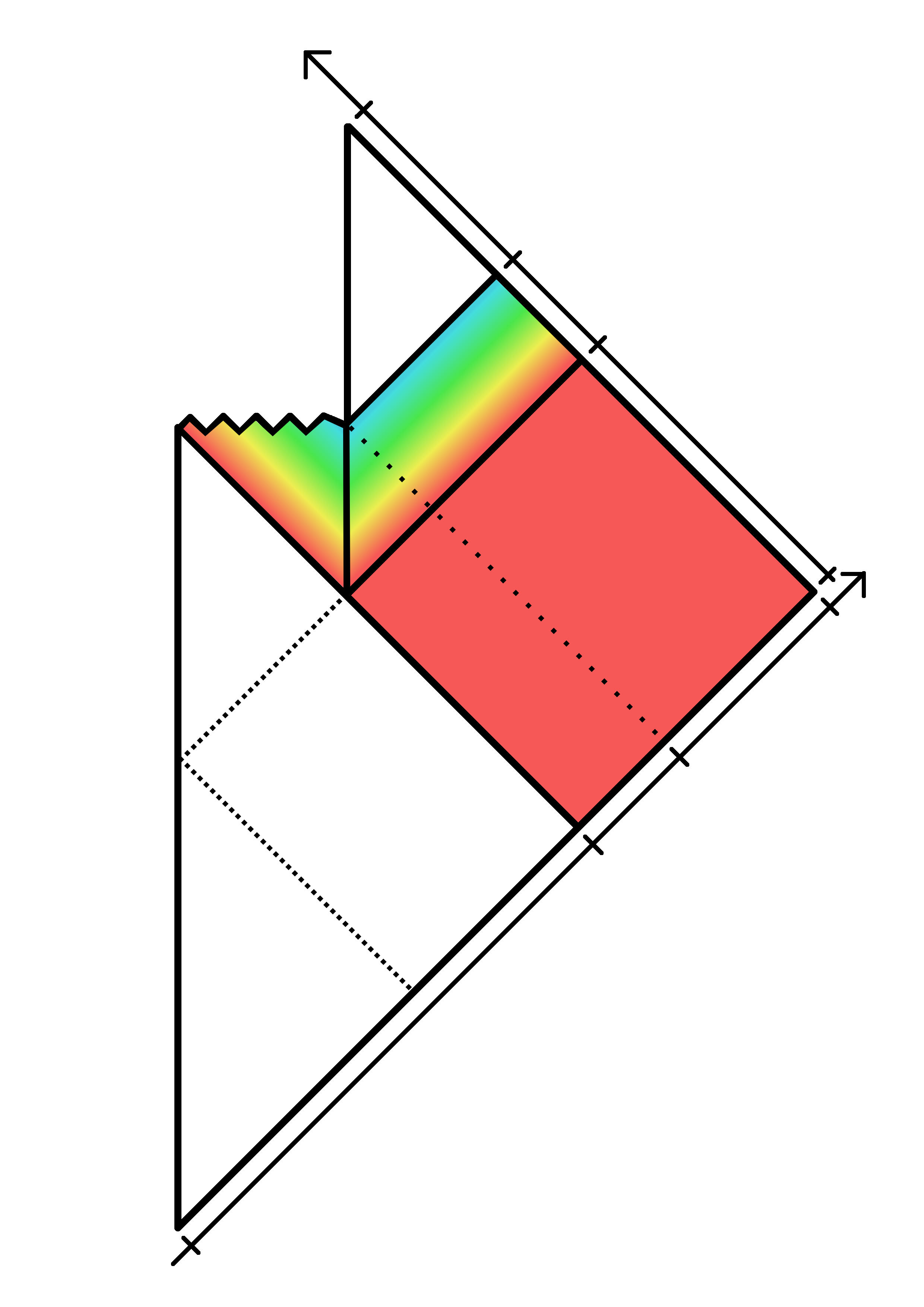}
\put (20,23) {$Ia$}
\put (30,37) {$Ib$}
\put (17,55) {$Ic$}
\put (45,55) {$II$}
\put (20,62) {$III$}
\put (33,67) {$IV$}
\put (30,78) {$V$}
\put (68,55) {$V$}
\put (65,50) {$\frac{\pi}{2}$}
\put (55,40) {$V_0 $}
\put (48,32) {$\frac{\pi}{4} $}
\put (18,3) {$\frac{\pi}{2} - 2 V_0 $}
\put (18,95) {$U$}
\put (30,93) {$0$}
\put (40,82) {$ V_0 - \frac{\pi}{2}$}
\put (48,75) {$- \frac{\pi}{4}$}
\put (61,60) {$- \frac{\pi}{2}$}
\end{overpic}
	\caption{}
	\label{HiscockII-subdivided}
\end{figure}

The Penrose diagram of Figure \ref{simple-black-hole} is a faithfull representation of the space-time described by the metric of equations (\ref{simple-black-hole-metricI}-\ref{simple-black-hole-metricIIb}). The explicit expression of the map that relates the coordinates of the diagram and that of the metric requires to subdivide the Penrose diagram, as shown on Figure \ref{HiscockII-subdivided}. 
Then it is given by the equations:
\begin{align}
\label{HiscockII-Ia}(Ia) &
\left[
\begin{array}{l}
u =  - 4m \left[ 1 +  W\left( - \frac{\tan U }{e} \right) \right] \\
v = - 4m \left[ 1 + W\left( - \frac{\tan \left( V + 2V_0 - \pi \right) }{e} \right) \right]
\end{array}
\right. \\
\label{HiscockII-Ib} (Ib) &
\left[
\begin{array}{l}
u =  - 4m \left[ 1 +  W\left( - \frac{\tan U }{e} \right) \right] \\
v = f_1(V) \ \text{increasing, such that} \\
 \left\{ \begin{array}{l}
f_1(-2V_0 + 3 \pi/4)= -4m (1+W(1/e))\\
f_1(\pi/4) = 0 
\end{array} \right.
\end{array}
\right. \\
(Ic) &
\left[
\begin{array}{l}
u = c_1 + f_1(U - 2V_0 + \pi) \\
v = c_1 + f_1(V)
\end{array}
\right. \\
(II) &
\left[
\begin{array}{l}
u = - 4m\log \left(-\tan U \right) \\
v = 4m \log \tan V 
\end{array}
\right. 
\end{align}
\begin{align}
\label{HiscockII-III} (III) &
\left[
\begin{array}{l}
v = f_2(V) \ \text{increasing, such that} \\
\hfill f_2(\pi/4) = N^{-1}(M(0)) \\
r = g(U,V) \ \text{such that} \\
 \left\{ \begin{array}{l}
\frac{\partial g}{\partial V}  = \frac{f_2'(V)}{2} \left( 1 - \frac{2 N(f_2(V))}{g(U,V)} \right) \\
\hfill g(U,\pi/4) = - \frac{1}{2}  f_1(U - 2V_0 + \pi) \\
g(2V_0 - \pi/2 - V,V) = 0 \\
\end{array} \right.
\end{array} 
\right. \hfill
\end{align}
\begin{align}
\label{HiscockII-IV}(IV) &
\left[
\begin{array}{l}
u = M^{-1}(N(f_2(U + \pi/2))) \\
r = h(U,V) \ \text{such that} \\
 \left\{ \begin{array}{l}
\frac{\partial h}{\partial U}  = - \frac{u'(U)}{2} \left( 1 - \frac{2 M(u(U))}{h(U,V)} \right) \\
h(-\pi/4,V) = 2m \left( 1 + W\left( \frac{ \tan V}{e} \right) \right) \\
h(U, \pi/2) = \infty \\
h(U,U+\pi/2) = g(U,U+\pi/2)
\end{array}
\right.
\end{array}
\right. \\
(V) &
\left[
\begin{array}{l}
v = M^{-1}(N(f_2(V_0)))  + 2 h (V_0-\pi/2,V) \\
u = M^{-1}(N(f_2(V_0))) + 2 h (V_0-\pi/2, U + \pi/2)
\end{array}
\right.
\end{align}
With these expressions we can check the consistency of the space-time model, and notably the gluing conditions, which match the metric along the boundaries of the patches. Moreover, the advanced time $v$ and the retarded time $u$ have been chosen to be both continuous along, respectively, $\mathcal{J}^-$ and $\mathcal{J}^+$ (this requirement is helpful to obtain the ray-tracing map). 

The metric depends on the parameters $m$ (the mass) and $V_0$ (linked to the life-time of the black hole) and an arbitrary constant $c_1$. Besides, the function $f_1, f_2, g, h, M, N$ are not given explicitely:
\begin{itemize}
\item $f_1$ and $f_2$ are arbitrary monotonous functions satisfying the boundary conditions given in  eq. \eqref{HiscockII-Ib} and eq. \eqref{HiscockII-III}.
\item $g$ and $h$ are fixed implicitely by  first order differential equations \eqref{HiscockII-III} and \eqref{HiscockII-IV}. These equations are obtained from the requirement that the lines of constant $V$ or $U$ are null. No explicit solution is known, except when $M$ and $N$ are constant or linear \cite{Waugh1986}.
\item $M$ and $N$ matches along the apparent horizon, between regions $III$ and $IV$: it is the first equation of \eqref{HiscockII-IV}. Thus, one of them can be freely chosen, depending on the expected phenomenology for the evaporation rate.
\end{itemize}

\section{Details of toy model}\label{detail toy model}

\begin{figure}[h]
\begin{overpic}[width =  0.9 \columnwidth]{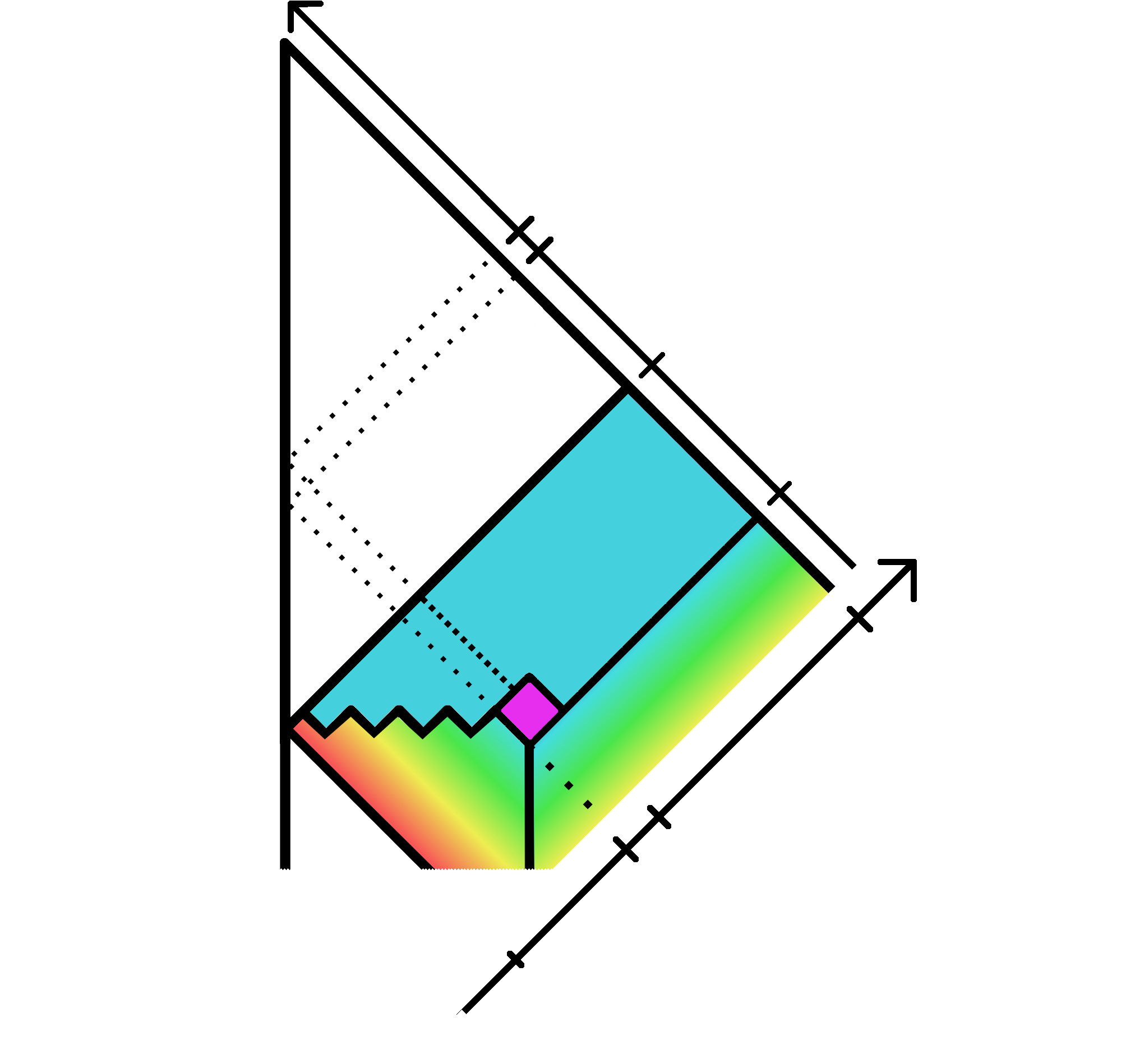}
\put (28,18) {$I$}
\put (38,22) {$III$}
\put (57,30) {$IV$}
\put (50,45) {$V$}
\put (10,35) {$VIa$}
\put(20,36){\vector(1,0){21}}
\put (32,32) {$VIb$}
\put (39,55) {$VIIa$}
\put (10,45) {$VIIb$}
\put(20,46){\vector(1,0){12}}
\put (10,40) {$VIIc$}
\put(20,41){\vector(1,0){10}}
\put (10,60) {$VIId$}
\put(20,61){\vector(1,0){15}}
\put (10,50) {$VIIe$}
\put(20,51){\vector(1,0){5}}
\put (29,70) {$VIIf$}
\put (85,40) {$V$}
\put (80,34) {$\frac{\pi}{2}$}
\put (58,14) {$V_0-\varepsilon $}
\put (50,4) {$\frac{\pi}{4} $}
\put (19,94) {$U$}
\put (60,65) {$ 2V_0 - \frac{3\pi}{4}$}
\put (71,54) {$ V_0 -\varepsilon- \frac{\pi}{2}$}
\end{overpic}
	\caption{Subdivided close-up of the Penrose diagram of a toy model of an evaporating black hole that turns into a white hole.}
	\label{close-up-model1}
\end{figure}
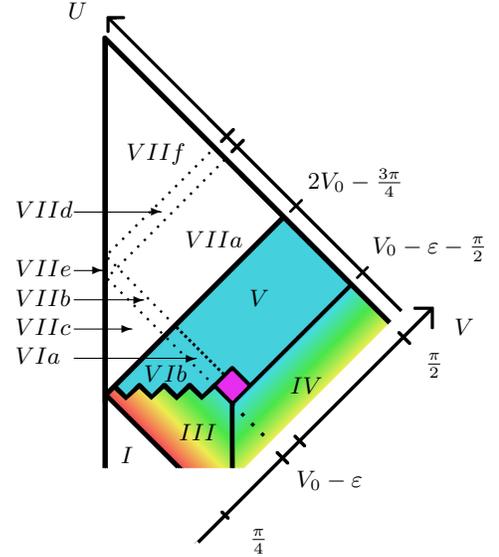

The coordinate map of the toy model in region $I-IV$ is the same as given by equations (\ref{HiscockII-Ia}-\ref{Hiscock metric IV}). Giving the coordinate map in the other regions require first to subdivide the Penrose diagram as is shown on Figure \ref{close-up-model1}.
The coordinates of the metric are related to that of the Penrose diagram through:
\begin{align}
(V) &
\left[
\begin{array}{l}
u = f_4(U) \ \text{increasing, such that} \\
\hfill f_4(V_0- \varepsilon-\pi/2)= M^{-1}(N(f_2(V_0-\varepsilon))) \\
v = f_4(V_0- \varepsilon-\pi/2) + 2 h(V_0- \varepsilon-\pi/2 , V ) \\
\hfill + 4m_1 \log \left( \frac{h(V_0- \varepsilon-\pi/2,V)}{2m_1} - 1 \right)
\end{array}
\right. \\
(VIa) &
\left[
\begin{array}{l}
u = f_5(U) \ \text{increasing} \\
v = - f_5(V_0- \varepsilon-\pi/2) -  2 h(V_0- \varepsilon-\pi/2 , V ) \\
\hfill - 4m_1 \log \left(1 - \frac{h(V_0- \varepsilon-\pi/2,V)}{2m_1} \right)
\end{array}
\right. \\
(VIb) &
\left[
\begin{array}{l}
u = c_5 + f_5(U) \\
v = - c_5 - f_5(2V_0 - \pi/2 - V)
\end{array}
\right. 
\end{align}
\begin{align}
(VIIa) &
\left[
\begin{array}{l}
u =  f_4(2V_0-3\pi/4) \\
+ 4m_1 \left( 1+ W(- e^{-\frac{f_5(2V_0-3\pi/4)-f_5(4V_0 - 3\pi/2 - U)}{4m_1}-1}) \right) \\
v = f_4(2V_0-3\pi/4) \\
\hfill + 4m_1 \left( 1+W((\frac{h(V_0- \varepsilon-\pi/2,V)}{2m_1}-1) \right. \\
\hfill \times \left. e^{\frac{f_4(V_0- \varepsilon-\pi/2)  - f_4(2V_0-3\pi/4)}{4m_1} + \frac{h(V_0- \varepsilon-\pi/2,V)}{2m_1} -1} ) \right) \\
\end{array}
\right. \\
(VIIb) &
\left[
\begin{array}{l}
u = c_6 + 4m_1  W\left(- e^{-\frac{f_5(2V_0-3\pi/4)-f_5(4V_0 - 3\pi/2 - U)}{4m_1}-1}\right)  \\
v = c_6 + 4m_1  W\left((\frac{h(V_0- \varepsilon-\pi/2,V)}{2m_1}-1) \right. \\
\times \left. e^{\frac{f_5(V_0- \varepsilon-\pi/2) - f_5(2V_0 - 3 \pi/4)}{4m_1} + \frac{h(V_0- \varepsilon - \pi/2 , V)}{2m_1}- 1} \right) \\
\end{array}
\right.
\end{align}
\begin{align}
(VIIc) &
\left[
\begin{array}{l}
u = c_7 + 4m_1 W\left(- e^{-\frac{f_5(2V_0-3\pi/4)-f_5(4V_0 - 3\pi/2 - U)}{4m_1}-1} \right)\\
v = c_7 + 4m_1   W\left(- e^{-\frac{f_5(2V_0-3\pi/4)-f_5(2V_0 - \pi/2 - V)}{4m_1}-1} \right)\\
\end{array}
\right. 
\end{align}
\begin{align}
(VIId) &
\left[
\begin{array}{l}
u =  f_4(2V_0 - 3\pi/4) \\
\hfill + 4m_1\left( 1 +  W((\frac{h(V_0- \varepsilon-\pi/2,U-2V_0+\pi)}{2m_1}-1) \right. \\
\hfill\times \left. e^{\frac{f_5(V_0- \varepsilon-\pi/2) - f_5(2V_0 - 3 \pi/4)}{4m_1} + \frac{h(V_0- \varepsilon - \pi/2 , U-2V_0+\pi)}{2m_1}- 1} \right)  \\
v = f_4(2V_0 - 3\pi/4) \\
\hfill + 4m_1  \left(1 + W((\frac{h(V_0- \varepsilon-\pi/2,V)}{2m_1}-1) \right. \\
\hfill\times \left. e^{\frac{f_4(V_0- \varepsilon-\pi/2)  - f_4(2V_0-3\pi/4)}{4m_1} + \frac{h(V_0- \varepsilon-\pi/2,V)}{2m_1} -1} ) \right) \\
\end{array}
\right. 
\end{align}
\begin{align}
(VIIe) &
\left[
\begin{array}{l}
u =  c_8 + 4m_1  W\left((\frac{h(V_0- \varepsilon-\pi/2,U-2V_0+\pi)}{2m_1}-1) \right. \\
\hfill\times \left. e^{\frac{f_5(V_0- \varepsilon-\pi/2) - f_5(2V_0 - 3 \pi/4)}{4m_1} + \frac{h(V_0- \varepsilon - \pi/2 , U-2V_0+\pi)}{2m_1}- 1} \right)  \\
v = c_8 + 4m_1  W\left((\frac{h(V_0- \varepsilon-\pi/2,V)}{2m_1}-1) \right. \\
\hfill\times \left. e^{\frac{f_5(V_0- \varepsilon-\pi/2) - f_5(2V_0 - 3 \pi/4)}{4m_1} + \frac{h(V_0- \varepsilon - \pi/2 , V)}{2m_1}- 1} \right) \\
\end{array}
\right. \\
(VIIf) &
\left[
\begin{array}{l}
u =  f_4(2V_0 - 3\pi/4) \\
\hfill + 4m_1  \big(1 +  W((\frac{h(V_0- \varepsilon-\pi/2,U-2V_0 + \pi)}{2m_1}-1) \\
\hfill\times \left. e^{\frac{f_4(V_0- \varepsilon-\pi/2)  - f_4(2V_0-3\pi/4)}{4m_1} + \frac{h(V_0- \varepsilon-\pi/2,U-2V_0 + \pi)}{2m_1} -1} ) \right) \\
v = f_4(2V_0 - 3\pi/4) \\
\hfill + 4m_1  \left(1 + W((\frac{h(V_0- \varepsilon-\pi/2,V)}{2m_1}-1) \right. \\
\hfill\times \left. e^{\frac{f_4(V_0- \varepsilon-\pi/2)  - f_4(2V_0-3\pi/4)}{4m_1} + \frac{h(V_0- \varepsilon-\pi/2,V)}{2m_1} -1} ) \right) \\
\end{array}
\right.
\end{align}

\section{Details of the evaporating black hole model with inside outgoing radiation}\label{detail inside outgoing}

\begin{figure}[h]
\begin{overpic}[width =  0.9 \columnwidth]{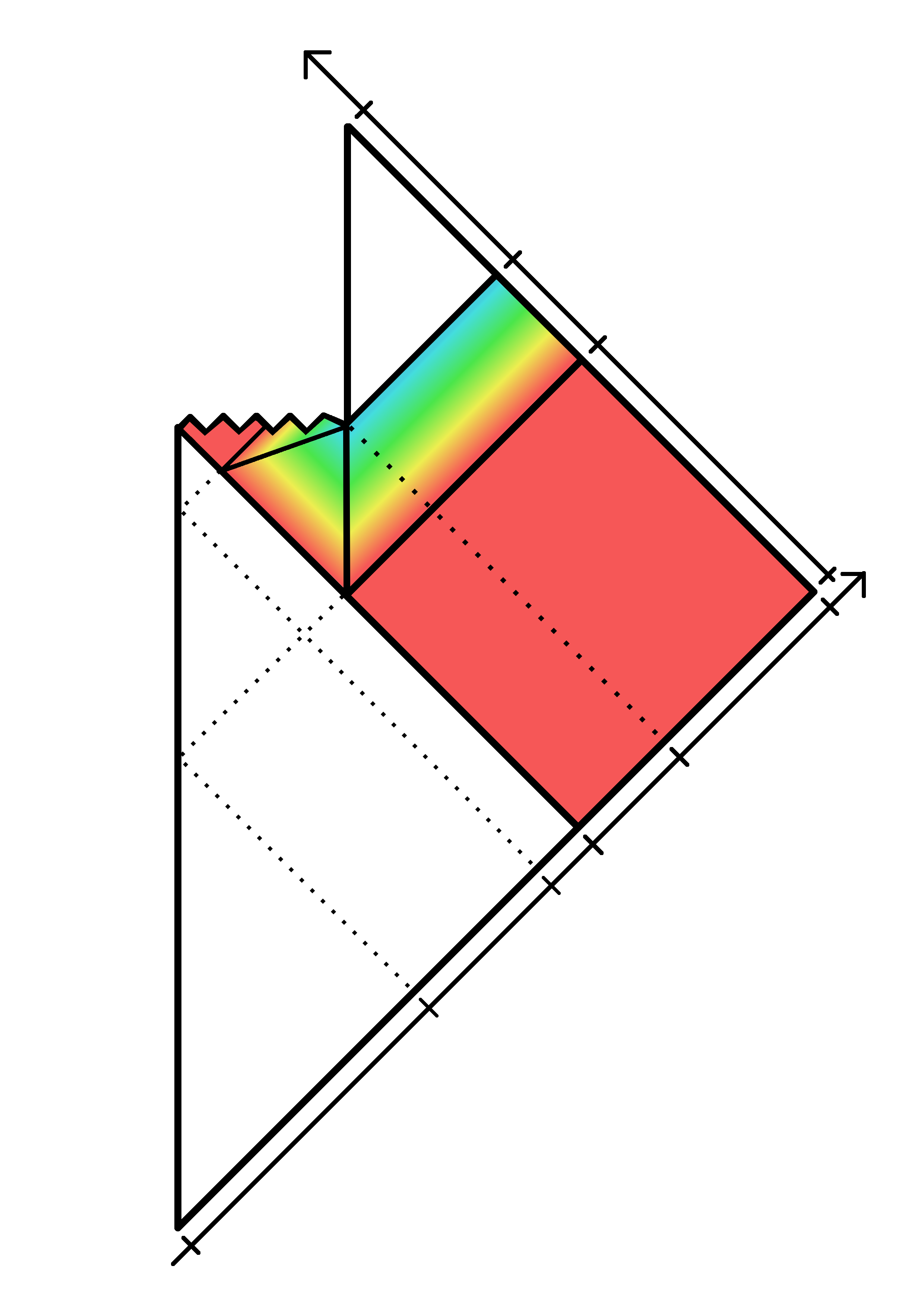}
\put (16,20) {$Ia$}
\put (25,35) {$Ib$}
\put (33,43) {$Ic$}
\put (16,50) {$Id$}
\put (20,55) {$Ie$}
\put (10,63) {$If$}
\put (45,55) {$II$}
\put (20,62) {$III$}
\put (33,67) {$IV$}
\put (30,78) {$V$}
\put (21,71) {$VI$}
\put(22,70){\vector(0,-1){3.8}}
\put (13,71) {$VII$}
\put(17,70){\vector(0,-1){4}}
\put (68,55) {$V$}
\put (65,50) {$\frac{\pi}{2}$}
\put (55,40) {$V_0 $}
\put (48,32) {$\frac{\pi}{4} $}
\put (43,27) {$V_1 $}
\put (34,17) {$\frac{3 \pi}{4} - 2V_0$}
\put (18,3) {$\frac{\pi}{2} - 2 V_0 $}
\put (18,95) {$U$}
\put (30,93) {$0$}
\put (40,82) {$ V_0 - \frac{\pi}{2}$}
\put (48,75) {$- \frac{\pi}{4}$}
\put (61,60) {$- \frac{\pi}{2}$}
\end{overpic}
	\caption{Subdivided Penrose diagram of an evaporating black hole with inside \textit{outgoing} flux.}
	\label{alterHiscock-subdivided}
\end{figure}

The subdivision of the Penrose diagram is on Figure \ref{alterHiscock-subdivided}.
\begin{align}
(Ia) &
\left[
\begin{array}{l}
u =  - 4m \left[ 1 +  W\left( - \frac{\tan U }{e} \right) \right] \\
v = - 4m \left[ 1 + W\left( - \frac{\tan \left( V + 2V_0 - \pi \right) }{e} \right) \right]
\end{array}
\right. \\
\label{alterHiscockII-Ib} (Ib) &
\left[
\begin{array}{l}
u =  - 4m \left[ 1 +  W\left( - \frac{\tan U }{e} \right) \right] \\
v = f_1(V) \ \text{increasing, such that}  \\
\hfill f_1(\frac{3 \pi}{4} - 2V_0) = -4m(1+W(1/e))
\end{array}
\right. \\
(Ic) &
\left[
\begin{array}{l}
u =  - 4m \left[ 1 +  W\left( - \frac{\tan U }{e} \right) \right]  \\
v = f_2(V) \ \text{increasing, such that}   \\
\left\{
\begin{array}{l}
f_2(V_1) = f_1(V_1)\\
f_2(\frac{\pi}{4})=0 
\end{array}
\right.
\end{array}
\right. \\
(Id) &
\left[
\begin{array}{l}
u =  c_1  + f_1(U-2V_0+\pi)  \\
v = c_1 + f_1(V) \\
\end{array}
\right. \\
(Ie) &
\left[
\begin{array}{l}
u =   c_2  + f_1(U-2V_0+\pi) \\
v = c_2  + f_2(V)  \\
\end{array}
\right. \\
(If) &
\left[
\begin{array}{l}
u =   c_3  + f_2(U-2V_0+\pi) \\
v = c_3  + f_2(V)  \\
\end{array}
\right. 
\end{align}
\begin{align}
(II) &
\left[
\begin{array}{l}
u = - 4m\log \left(-\tan U \right) \\
v = 4m \log \tan V 
\end{array}
\right. \\
\label{alterHiscockII-III} (III) &
\left[
\begin{array}{l}
v = f_3(V) \ \text{monotonous, such that} \\
\hfill f_3(\pi/4) = N^{-1} (M(0)) \\
r = g(U,V) \ \text{such that} \\
 \left\{ \begin{array}{l}
\frac{\partial g}{\partial V}  = \frac{f_3'(V)}{2} \left( 1 - \frac{2 N(f_3(V))}{g(U,V)} \right) \\
\hfill g(U,\pi/4) = -  \frac{1}{2} f_1(U - 2V_0 + \pi) 
\end{array} \right.
\end{array}
\right. 
\end{align}
\begin{align}
\label{alterHiscockII-IV}(IV) &
\left[
\begin{array}{l}
u = M^{-1}(N(f_3(U + \pi/2))) \\
r = h(U,V) \ \text{such that} \\
 \left\{ \begin{array}{l}
\frac{\partial h}{\partial U}  = - \frac{u'(U)}{2} \left( 1 - \frac{2 M(u(U))}{h(U,V)} \right) \\
h(-\pi/4,V) = 2m \left( 1 + W\left( \frac{ \tan V}{e} \right) \right) \\
h(U, \pi/2) = \infty \\
h(U,U+\pi/2)=g(U,U+\pi/2)
\end{array}
\right.
\end{array}
\right. \\
(V) &
\left[
\begin{array}{l}
v =  M^{-1}(N(f_3(V_0))) + 2 h (V_0-\pi/2,V) \\
u = M^{-1}(N(f_3(V_0))) + 2 h (V_0-\pi/2, U + \pi/2)
\end{array}
\right. 
\end{align}
\begin{align}
\label{alterHiscockII-VI}(VI) &
\left[
\begin{array}{l}
u = P^{-1}(N(f_3(\mathcal{C}^{-1}(U)))) \\
r = j(U,V) \ \text{such that} \\
 \left\{ \begin{array}{l}
\frac{\partial j}{\partial U}  = - \frac{u'(U)}{2} \left( 1 - \frac{2 P(u(U))}{j(U,V)} \right) \\
j(2V_0 - \pi + V_1,V) =  2m \left( 1 \right. \\
\hfill  \quad  \left. + W( - \frac{4m+f_2(V_1)}{4m + f_2(\pi/2- V)} e^{-\frac{f_2(V_1)}{4m} + \frac{f_2(\pi/2-V)}{4m} - 1} ) \right) \\
j(2V_0-\pi/2 - V,V) = 0 \\
j(\mathcal{C}(V),V) = g(\mathcal{C}(V),V)
\end{array}
\right.
\end{array}
\right. \\
\label{alterHiscockII-VII}(VII) &
\left[
\begin{array}{l}
u = - c_4 - f_2(U-2V_0+\pi) \\
\hfill + 4m \log \left( 1 + \frac{f_2(U-2V_0 + \pi)}{4m} \right)  \\
v = c_4 +f_2(\pi/2 - V) \\
\hfill - 4m \log \left( 1 + \frac{f_2(\pi/2 - V)}{4m} \right) 
\end{array}
\right.
\end{align}
There are three parameters $m$, $V_0$ and $V_1$. The constants $c_1, c_2, c_3$ and $c_4$ are arbitrary. The functions $g, h$ and $j$ are fixed implicitly by the differential equations. the functions $f_1, f_2$ and $f_3$ are arbitrary. The functions $P, M, N$ encode the phenomenology of the evaporation (two fix the position of the two pseudo-horizons, while the third fixes the rate of evaporation). The curve $\mathcal{C}$ parametrise the boundary $III/VI$ in the Penrose coordinates (it is space-like).

\vfill

\bibliographystyle{utcaps}
\bibliography{hawking}

\end{document}